\documentclass[longauth]{aa} 

\usepackage{graphicx}
\usepackage{txfonts}

\usepackage[colorlinks=true, citecolor=cyan,
 linkcolor=cyan, menucolor=blue, urlcolor=blue,
 linkbordercolor={0 0 1}, frenchlinks=True, breaklinks]{hyperref}
\usepackage{threeparttable}

\usepackage{color}

\begin{document} 

   \title{Physical Properties of the Southwest Outflow Streamer in the Starburst Galaxy NGC~253 with ALCHEMI}
   \titlerunning{Properties of Southwest Streamer in NGC~253}

   \author{Min Bao
          \inst{1,2,3}, 
          Nanase Harada
          \inst{4,5},
          Kotaro Kohno
          \inst{1},
          Yuki Yoshimura
          \inst{1},
          Fumi Egusa
          \inst{1},
          Yuri Nishimura
          \inst{1,6},
          Kunihiko Tanaka
          \inst{7},
          Kouichiro Nakanishi
          \inst{4,5},
          Sergio Mart\'in
          \inst{8,9},
          Jeffrey G. Mangum
          \inst{10},
          Kazushi Sakamoto
          \inst{11},
          S\'ebastien Muller
          \inst{12},
           Mathilde Bouvier
           \inst{13},
          Laura Colzi
          \inst{14},
          Kimberly L. Emig
          \inst{10},
          David S. Meier
          \inst{15,16},
          Christian Henkel
          \inst{17,18},
          Pedro Humire
          \inst{19,17},
          Ko-Yun Huang
          \inst{13},
          V\'ictor M. Rivilla
          \inst{14},
          Paul van der Werf
          \inst{13},
          Serena Viti
          \inst{13}
          }

   \institute{ 
            Institute of Astronomy, Graduate School of Science, The University of Tokyo, 2-21-1 Osawa, Mitaka, Tokyo 181-0015, Japan\\
            \email{kkohno@ioa.s.u-tokyo.ac.jp }
        \and 
            School of Astronomy and Space Science, Nanjing University, Nanjing 210023, China
        \and 
            School of Physics and Technology, Nanjing Normal University, Nanjing 210023, China
        \and 
            National Astronomical Observatory of Japan, 2-21-1 Osawa, Mitaka, Tokyo 181-8588, Japan\\
             \email{nanase.harada@nao.ac.jp}
        \and 
            Department of Astronomy, School of Science, The Graduate University for Advanced Studies (SOKENDAI), 2-21-1 Osawa, Mitaka Tokyo, 181-1855 Japan
        \and 
            ALMA Project, National Astronomical Observatory of Japan, 2-21-1, Osawa, Mitaka, Tokyo 181-8588, Japan
        \and 
            Department of Physics, Faculty of Science and Technology, Keio University, 3-14-1 Hiyoshi, Yokohama, Kanagawa 223–8522 Japan
        \and 
            European Southern Observatory, Alonso de C\'ordova, 3107, Vitacura, Santiago 763-0355, Chile
        \and 
            Joint ALMA Observatory, Alonso de C\'ordova, 3107, Vitacura, Santiago 763-0355, Chile
        \and 
            National Radio Astronomy Observatory, 520 Edgemont Road, Charlottesville, VA 22903-2475, USA
        \and 
            Institute of Astronomy and Astrophysics, Academia Sinica, 11F of AS/NTU Astronomy-Mathematics Building, No.1, Sec. 4, Roosevelt Rd, Taipei 10617, Taiwan
        \and 
            Department of Space, Earth and Environment, Chalmers University of Technology, Onsala Space Observatory, SE-439 92 Onsala, Sweden
        \and 
            Leiden Observatory, Leiden University, P.O. Box 9513, 2300 RA Leiden, The Netherlands
        \and 
            Centro de Astrobiolog\'ia (CSIC-INTA), Ctra. de Ajalvir Km. 4, Torrej\'on de Ardoz, 28850 Madrid, Spain
        \and 
            New Mexico Institute of Mining and Technology, 801 Leroy Place, Socorro, NM 87801, USA
        \and 
            National Radio Astronomy Observatory, PO Box O, 1003 Lopezville Road, Socorro, NM 87801, USA
        \and 
            Max-Planck-Institut f\"ur Radioastronomie, Auf dem H\"ugel 69, 53121 Bonn, Germany
        \and 
            Astronomy Department, Faculty of Science, King Abdulaziz University, P.~O.~Box 80203, Jeddah 21589, Saudi Arabia
        \and 
            Departamento de Astronomia, Instituto de Astronomia, Geofísica e Ciências Atmosféricas da USP, Cidade Universitária, 05508-900 São Paulo, SP, Brazil
             }

\authorrunning{Bao et al.}


  \abstract
   {}
   {The physical properties of galactic molecular outflows are important as they could constrain outflow formation mechanisms. In this work, we study the properties of the southwest (SW) outflow streamer including gas kinematics, optical depth, dense gas fraction, and shock strength through molecular emission in the central molecular zone of the starburst galaxy NGC~253.}
   {We image the molecular emission in NGC~253 at a spatial resolution of 1.6$^{\prime\prime}$($\sim$27\,pc at D$\sim$3.5\,Mpc) based on data from the ALMA Comprehensive High-resolution Extragalactic Molecular Inventory (ALCHEMI) large program. We trace the velocity and the velocity dispersion of molecular gas with CO(1-0) line and study the molecular spectra in the SW streamer region, the brightest CO streamer in NGC~253. We constrain the optical depth of CO emission with CO/$^{13}$CO(1-0) ratio, the dense gas fraction with HCN/CO(1-0), H$^{13}$CN/$^{13}$CO(1-0) and N$_{2}$H$^{+}$/$^{13}$CO(1-0) ratios, as well as the shock strength with SiO(2-1)/$^{13}$CO(1-0) and CH$_{3}$OH(2$_{k}$-1$_{k}$)/$^{13}$CO(1-0) ratios.}
   {The CO/$^{13}$CO(1-0) integrated intensity ratio is $\sim$21 in the SW streamer region, which approximates the C/$^{13}$C isotopic abundance ratio. The higher integrated intensity ratio compared to the disk can be attributed to the optically thinner environment for CO(1-0) emission inside the SW streamer. The HCN/CO(1-0) and SiO(2-1)/$^{13}$CO(1-0) integrated intensity ratios both approach $\sim$0.2 in three giant molecular clouds (GMCs) at the base of the outflow streamers, which implies the higher dense gas fraction and enhanced strength of fast shocks in those GMCs than in the disk, while the HCN/CO(1-0) integrated intensity ratio is moderate in the SW streamer region. The contours of those two integrated intensity ratios are extended towards the directions of outflow streamers, which connects the enhanced dense gas fraction and shock strength with molecular outflow. Moreover, the molecular gas with enhanced dense gas fraction and shock strength located at the base of the SW streamer shares the same velocity with the outflow.}
   {The enhanced dense gas fraction and shock strength at the base of the outflow streamers suggest that the star formation inside the GMCs can trigger the shocks and further drive the molecular outflow. The increased CO/$^{13}$CO(1-0) integrated intensity ratio coupled with the moderate HCN/CO(1-0) integrated intensity ratio in the SW streamer region are consistent with the picture that the gas velocity gradient inside the streamer may decrease the optical depth of CO(1-0) emission, as well as the dense gas fraction in the extended streamer region.}

   \keywords{Galaxies: individual: NGC~253 -- Galaxies: evolution --
             Galaxies: starburst --
            Galaxies: kinematics and dynamics}

   \maketitle

\section{Introduction}
\label{introduction}

Outflows on the galactic scale are ubiquitous in the local and distant Universe. Together with gas accretion, star formation, and black hole growth, outflows can govern the cycle of material between a galaxy and the circumgalactic medium \citep{2005ARA&A..43..769V}. Cool outflows, which include atomic and molecular gas, as well as dust, dominate the mass and energy of the outflowing material. The study of extragalactic cool outflows only dates back to $\sim$20 years, which is a relatively new field in the research of galactic outflows \citep{2020A&ARv..28....2V}. The physical properties of cool outflows are gradually uncovered thanks to the development of high-capability radio telescopes, including ALMA \citep{2015ApJ...808L...1A}, NOEMA \citep{2020EPJWC.22800014L}, and VLA \citep{2003ASSL..285..109G}.

As one of the key phases of cool gas, molecular gas is the raw material for star formation, an important process in galaxy evolution. More than 300 molecular species have been detected in the circumstellar envelopes of evolved stars and the interstellar medium (ISM) of the Galaxy \footnote{\url{https://cdms.astro.uni-koeln.de/classic/molecules}}. One-third of those species have also been detected in external galaxies \citep{2019PASJ...71S..20T, 2021A&A...656A..46M}. Among them, carbon monoxide (CO) is the second most abundant molecule after H$_{2}$ and serves as the main observational tracer of molecular gas. Moreover, other molecules with different chemical formation pathways and excitation requirements provide supplementary information to constrain the evolution of external galaxies \citep{2015ASPC..499...85A}.

The formation scenarios for molecular outflows can be roughly divided into three categories. The outflowing molecular gas can be directly driven by radiation \citep{2011ApJ...735...66M} and/or pressure gradients \citep{2008ApJ...687..202S, 2012MNRAS.423.2374U}. Alternatively, hot winds can entrain molecular clouds \citep{2019MNRAS.486.4526B, 2022ApJ...924...82F}, while the lifetime of the molecular clouds depends on the balance between radiation, conduction, and turbulence \citep{2005A&A...444..505O}. The third scenario is that molecular gas forms in situ from hot winds through cooling and/or thermal instability. The first step in the formation of molecules consists of hydrogen atoms (H) combining to form hydrogen molecules (H$_{2}$). Given that H atoms can combine on the surface of dust grains, those grains are catalysts for the formation of H$_{2}$ \citep{1971ApJ...163..155H, 2021ApJ...917...49P}. However, along with temperature rises, dust grains undergo destructive sputtering via gas-grain collisions in the hot winds, which reduces the efficiency of H$_{2}$ formation \citep{2012A&A...541A..76L}. Hence, a cooling timescale shorter than a dynamical timescale is a premise for the in situ formation \citep{2000MNRAS.317..697E, 2003ApJ...590..791S}.

There is no agreement on the influence of outflows on the star formation activity inside host galaxies. On the one hand, the hot ionized winds are one of the leading processes that can quench star formation \citep{2015Sci...348..314T, 2016MNRAS.458..242T, 2019ApJ...883...81S}, namely, negative feedback. The quenching is owing to the energy that is injected into molecular clouds, or to the entrainment/depletion of molecular clouds. On the other hand, the shocks, which form through the encounter between winds and molecular clouds, can compress the molecular clouds and trigger star formation \citep{1994ApJ...420..213K}, namely, positive feedback.

The engines for outflows can be central starbursts and/or active galactic nuclei (AGN). Different physical processes including thermal energy, radiation, cosmic rays, and/or radio jets can work together in either driving mechanism. AGN-driven molecular outflows have been well studied in a few nearby cases such as the Seyfert 2 galaxy NGC 1068 \citep{2013ARA&A..51..511K} and the Quasar Mrk 231 \citep{2009ApJ...701..587V}. However, to study how molecular outflows relate to star formation, we need to target sources without contamination from AGN.

NGC~253 (D$\sim$3.5\,Mpc, \citealt{2005MNRAS.361..330R}) is one of the nearest starburst galaxies. It possesses a strong bar in the center \citep{2001ApJ...549..896D, 2004ApJ...611..835P}, but shows no sign of AGN activity \citep{2009A&A...497..103B}. It is an edge-on (i$\sim$78.5$^{\circ}$) spiral galaxy and hosts a bipolar outflow \citep{1985ApJ...299..312T, 2011MNRAS.414.3719W}. The outflow was proven to be driven by the starburst that has been occurring in the central $\sim$500 pc region during the last $\sim$20-30\,Myr \citep{1980ApJ...238...24R, 1998ApJ...505..639E}, where the star formation rate is $\sim$3\,M$_{\odot}$\,yr$^{-1}$ \citep{2005ApJ...629..767O, 2015MNRAS.450L..80B}. The outflow has been detected in multi-phases, including molecular gas phase \citep{1985ApJ...299..312T, 2011ApJ...733L..16S, 2011A&A...535A..79H, 2013Natur.499..450B, 2017ApJ...835..265W, 2018ApJ...867..111Z, 2019ApJ...881...43K}, ionized gas phase \citep{2000ApJS..129..493H, 2011MNRAS.414.3719W, 2020MNRAS.493..627C}, X-ray emitting gas phase \citep{2000A&A...360...24P, 2000AJ....120.2965S, 2002ApJ...568..689S, 2023ApJ...942..108L}, and dust phase \citep{2022ApJ...935...19L}. The ionized outflow shows a wide opening angle of $\sim$60$^{\circ}$, a deprojected velocity of a few 100\,km\,s$^{-1}$ \citep{2011MNRAS.414.3719W}, and an extension of $\sim$10\,kpc \citep{2002ApJ...568..689S}.

There have been several ALMA-based studies of the molecular outflow of NGC~253. \cite{2019ApJ...881...43K} presented ALMA observation of CO(3-2) in the central 30$^{\prime\prime}$ starburst region. They obtained the non-disk component by modeling the disk, deducted the disk from position-velocity diagrams (PVDs), and found that $\sim$7\%-16\% of CO luminosity is emitted by the non-disk component. \cite{2013Natur.499..450B} presented ALMA observation of CO(1-0) in the central arcminute, and found the extraplanar molecular gas closely tracking the H$\alpha$ filaments; they estimated the mass outflow rate assuming an optically thin conversion factor ($\alpha_{\rm CO}$). By comparing the molecular mass outflow rate (9\,$M_{\odot}$ yr$^{-1}$) with the star formation rate \citep[2.8\,$M_{\odot}$ yr$^{-1}$,][]{2005ApJ...629..767O}, \cite{2013Natur.499..450B} concluded that the starburst-driven outflow limits the star-formation activity in NGC~253. \cite{2018ApJ...867..111Z} presented ALMA observations of CO(1-0) and CO(2-1) in the central 40$^{\prime\prime}$ region. They located the molecular outflow via the extended structure in the velocity-integrated intensity map of CO(2-1). Based on the velocity-integrated intensity ratio between CO(2-1) and CO(1-0), they constrained the optical depth of the CO emission in the outflow region. \cite{2017ApJ...835..265W} presented a detailed study of one streamer of the molecular outflow on the southwestern side (SW streamer). In addition to the CO emission, they found many bright tracers of dense gas such as HCN, CN, HCO$^{+}$, and CS in the SW streamer.

This paper is based on the data from the ALMA Comprehensive High-resolution Extragalactic Molecular Inventory (ALCHEMI) project \citep{2021A&A...656A..46M}. The ALCHEMI data targeting the central molecular zone (CMZ) of NGC~253 with high resolution, provide insight into the physical condition in the center of this galaxy \citep{2022A&A...659A.158H}. Those data have been used to study the abundance and excitation of different molecular species \citep{2021A&A...654A..55H, 2022ApJ...938...80H} in the CMZ of NGC~253, and how they vary with different heating environments \citep{2021ApJ...923...24H, 2022ApJ...931...89H, 2022ApJ...939..119B}. In this paper, we uncover the physical properties of the molecular outflow in NGC~253, focusing on the properties of the SW streamer. Taking advantage of the numerous molecular species in the ALCHEMI survey, we study the properties through various integrated intensity ratios. In detail, we use the CO/$^{13}$CO(1-0) ratio to probe the optical depth \citep{2020A&A...635A.131I}, HCN/CO(1-0), H$^{13}$CN/$^{13}$CO(1-0) and N$_{2}$H$^{+}$/$^{13}$CO(1-0) ratios to probe the dense gas fraction \citep{2004ApJ...606..271G, 2020MNRAS.497.1972B}. Moreover, we use SiO(2-1)/$^{13}$CO(1-0) and CH$_{3}$OH(2$_{k}$-1$_{k}$)/$^{13}$CO(1-0) ratios to probe the shock strength \citep{2010A&A...519A...2G}.

This paper is organized as follows. The data analysis is presented in Section \ref{data analysis}. The physical properties of the molecular outflow, including gas kinematics, optical depth, dense gas fraction, and shock strength, are studied in Section \ref{physical properties of the molecular outflow}. In Section \ref{formation of molecular outflow}, we analyse the correlation between the dense gas fraction and the strength of fast shocks, and discuss how the molecular outflow relates to the star formation in NGC~253. Finally, the results are summarized in Section \ref{summary}.

\section{Data analysis}
\label{data analysis}

The data used in this study have been obtained as part of the ALCHEMI survey, which is an ALMA Cycle 5 large program (2017.1.00161.L). It consists of a wide and unbiased spectral survey covering the CMZ of NGC~253 in Bands 3, 4, 6, and 7. The phase center of observation is $\alpha = 00^{\rm h}47^{\rm m}33.26^{\rm s}$, $\delta = -25^{\circ}17^{\prime}17.7^{\prime\prime}$. In this paper, we only make use of the Band 3 data, which covers the ground-state transitions of common molecular lines that can be emitted even from the coldest ($\sim$10\,K) medium. We refer readers to \cite{2021A&A...656A..46M} for more details about the data analysis, and only enumerate here the relevant information to this work.

The data cubes from the ALCHEMI survey are imaged to a spatial resolution of 1.6$^{\prime\prime}$  ($\sim27$\,pc) and a spectral resolution of $\Delta$v$\sim$10\,km\,s$^{-1}$. We uniformly use the continuum emission subtracted cubes, which are created by the Python-based tool \texttt{STATCONT} \citep{2018A&A...609A.101S}. Those cubes are primary beam corrected in the ALCHEMI imaging process. The H\"ogbom deconvolver function is used for all the cubes from the ALCHEMI survey. Meanwhile, we produce the self-calibrated data cubes with a multi-scale deconvolver function for the integrated intensities of CO and $^{13}$CO in the J = 1-0 transition in Figs. \ref{fig5}(a) and \ref{fig10}(a). The use of self-calibration allows for a higher signal-to-noise ratio and is better for images with complicated spatial structures, which helps us in precisely obtaining the optical depth of CO emission in both disk and outflow regions (see Section \ref{sec:opt_depth}).

The data cubes from the ALCHEMI survey are subject to the missing flux problem due to the lack of short spacing in interferometric observations. However, the band 3 observations here probe scales up to $\sim$30$^{\prime\prime}$, while spatial filtering should be relevant to larger scales \citep{2021A&A...656A..46M}. In order to quantify the missing fluxes, we collect the results of IRAM 30m single-dish observations from \cite{2015A&A...579A.101A}. We unify the beam size of the ALCHEMI data cubes to that of the IRAM 30m observations using the CASA command \texttt{imsmooth} \citep{CASATeam2022PASP} and compare the integrated intensities between them. All the missing fluxes of the molecular lines in this work are less than a few percent.

To ensure sufficient signal-to-noise ratio (S/N), we mask the region without 3$\sigma$-detection. Details of our procedures are as follows. We take the data cubes before primary beam correction as references, which share the same dimensions as those after primary beam correction. For each targeted molecular line, we mask the region without 3$\sigma$-detection for each channel in the corresponding data cube. The integrated intensity, velocity, and velocity dispersion maps, i.e., moments 0, 1, and 2 maps, in the CMZ (inner $\sim$500\,pc) of NGC~253 are generated by the CASA command \texttt{immoments} \citep{CASATeam2022PASP}. The PVDs are generated by the Python-based tool \texttt{pvextractor} \citep{2016ascl.soft08010G}. Given the 3$\sigma$-detection limits prior to analyses, all the features presented in the following maps and PVDs will be robust.

\section{Physical properties of the molecular outflow}
\label{physical properties of the molecular outflow}

\subsection{Molecular emission}

\begin{figure*}
   \centering
   \includegraphics[width=\hsize]{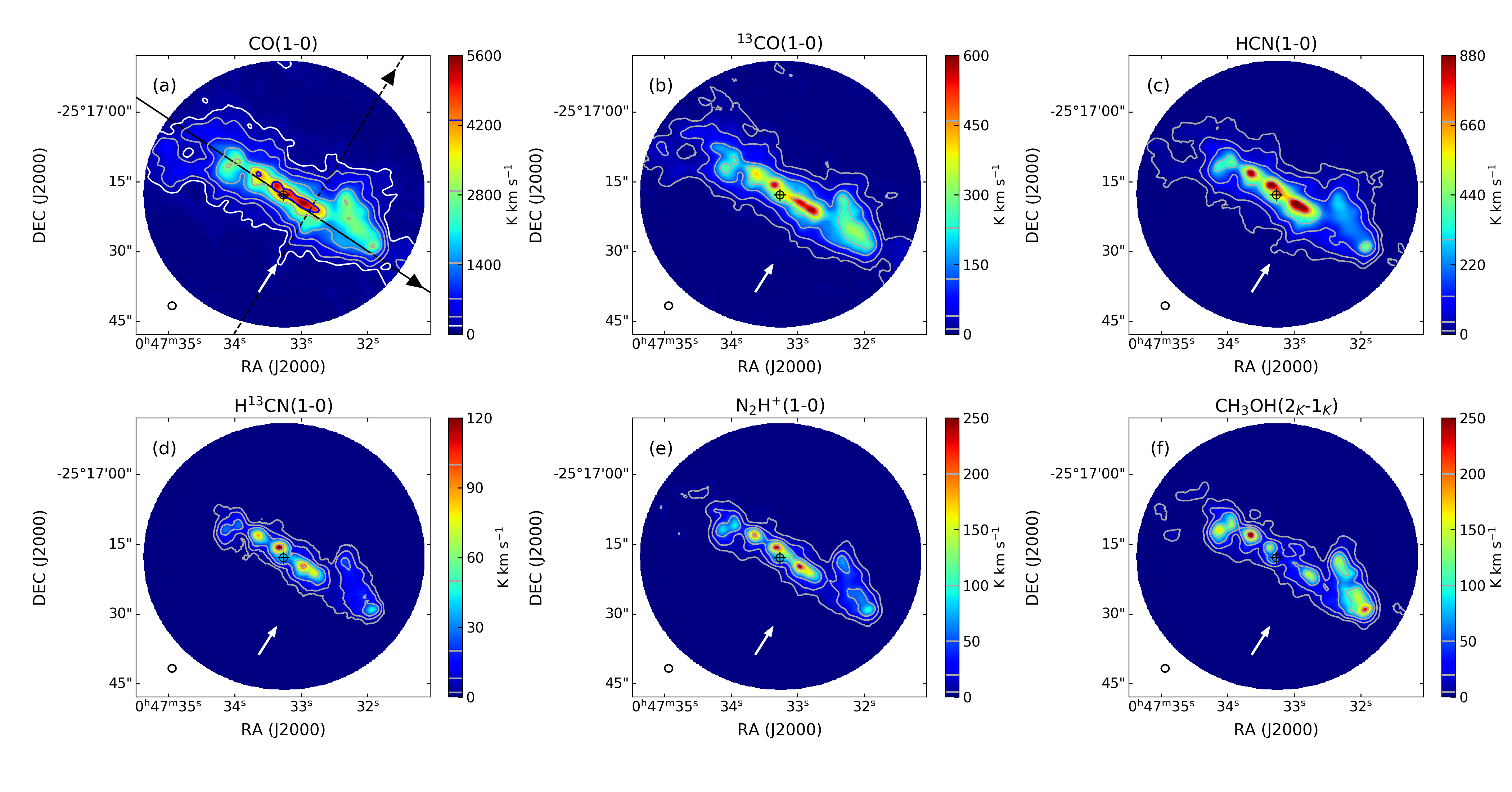}
   \caption{Integrated intensity maps of the transitions used in this work. (a) Integrated intensity map of CO(1-0). The white contour is drawn at 180\,K\,km\,s$^{-1}$, the grey contours are drawn at [360, 720, 1440, 2880]\,K\,km\,s$^{-1}$, and the blue contour is drawn at 4300\,K\,km\,s$^{-1}$. The horizontal lines in different colors inside the colorbar mark the values for corresponding contours, and remain the same in the following figures. black solid line marks the major axis, and the black dashed line marks the slice across the SW streamer, on which the black arrows show the positive directions for the PVDs in the following figures. The white arrow points to the SW streamer, and remains the same in the following panels. The black cross marks the phase center of the observation, and remains the same in the following figures. The beam size of 1.6$^{\prime\prime}$ is shown as a black empty circle at the bottom-left corner of each panel, and remains the same in the following figures. (b) Integrated intensity map of $^{13}$CO(1-0). The grey contours are drawn at [12, 40, 120, 230, 460]\,K\,km\,s$^{-1}$. (c) Integrated intensity map of HCN(1-0). The grey contours are drawn at [12, 40, 120, 300, 670]\,K\,km\,s$^{-1}$. (d) Integrated intensity map of H$^{13}$CN(1-0). The grey contours are drawn at [2, 8, 20, 50, 100]\,K\,km\,s$^{-1}$. (e) Integrated intensity map of N$_{2}$H$^{+}$(1-0). The grey contours are drawn at [5, 20, 50, 100, 200]\,K\,km\,s$^{-1}$. (f) Integrated intensity map of CH$_{3}$OH(2$_{k}$-1$_{k}$). The grey contours are drawn at [5, 20, 50, 100, 200]\,K\,km\,s$^{-1}$.}
   \label{fig1}
\end{figure*}

Fig. \ref{fig1} shows the integrated intensity maps of CO, $^{13}$CO, HCN, H$^{13}$CN, and N$_{2}$H$^{+}$ in the J = 1-0 transition and CH$_{3}$OH (methanol) in the J$_{k}$ =  2$_{k}$-1$_{k}$ transition series. We extract spectral channels covering $1000\,\rm km\,s^{-1}$ velocity range around each transition from the continuum-subtracted cubes, where the rest-frame frequencies are 115.271, 110.201, 88.632, 86.340, 93.174, and $\sim$96.741 GHz, respectively. The black solid line in Fig. \ref{fig1}(a) marks the galactic major axis (hereafter major axis) with a position angle of 55$^{\circ}$, and is defined following \cite{2019ApJ...881...43K}. The strong CO(1-0) emission in Fig. \ref{fig1}(a), which is marked by the blue contour at 4300\,K\,km\,s$^{-1}$, concentrates in the galactic center and distributes along the major axis. Similar structures also exist in other panels of Fig. \ref{fig1}. \cite{2001ApJ...549..896D} found that a bar structure along the major axis is necessary for modeling the ionized gas velocity field in the central $\sim$100\,pc region of NGC~253. A bar structure is also applicable in the CO map in the central $\sim$300\,pc region of NGC~253 \citep{2004ApJ...611..835P}. A stellar bar, which is known as an efficient mechanism for gas accretion, can trigger strong molecular emission along the major axis as is seen in Fig. \ref{fig1}. One different point in Fig. \ref{fig1}(f) from the other panels in Fig. \ref{fig1} is that the strongest CH$_{3}$OH(2$_{k}$-1$_{k}$) emission exists on the outskirts of the gas disk, suggesting quasi-thermal emission in agreement with \citet{2022A&A...663A..33H}.

The CO(1-0) contours (especially the white one at 180\,K\,km\,s$^{-1}$) in Fig. \ref{fig1}(a) are extended towards the SW streamer defined by \citet{2017ApJ...835..265W}, which is indicated by a white arrow. Similar extent can be found in the HCN(1-0) contours (Fig. \ref{fig1}c), while the $^{13}$CO(1-0) contours are not so extended (Fig. \ref{fig1}b). Based on the ratios between the main and rarer isotopologues, \cite{2015ApJ...801...63M} found that CO(1-0) and HCN(1-0) emission is on average optically thick in the central kpc-scale region of NGC~253. As $^{13}$C-bearing molecular species, $^{13}$CO(1-0) emission should be moderately optically thin \citep{2019A&A...624A.125M}. The different extents towards the SW streamer of CO(1-0) and HCN(1-0) emission from the $^{13}$CO(1-0) emission imply that the optical depths of CO(1-0) and HCN(1-0) emission in the SW streamer region may be lower than that in the gas disk. Moreover, the fainter emission of H$^{13}$CN(1-0), N$_{2}$H$^{+}$(1-0) and CH$_{3}$OH(2$_{k}$-1$_{k}$) in Figs. \ref{fig1}(d), \ref{fig1}(e) and \ref{fig1}(f) share similar extents towards the SW streamer with $^{13}$CO(1-0).

\begin{figure*}
   \centering
   \includegraphics[width=\hsize]{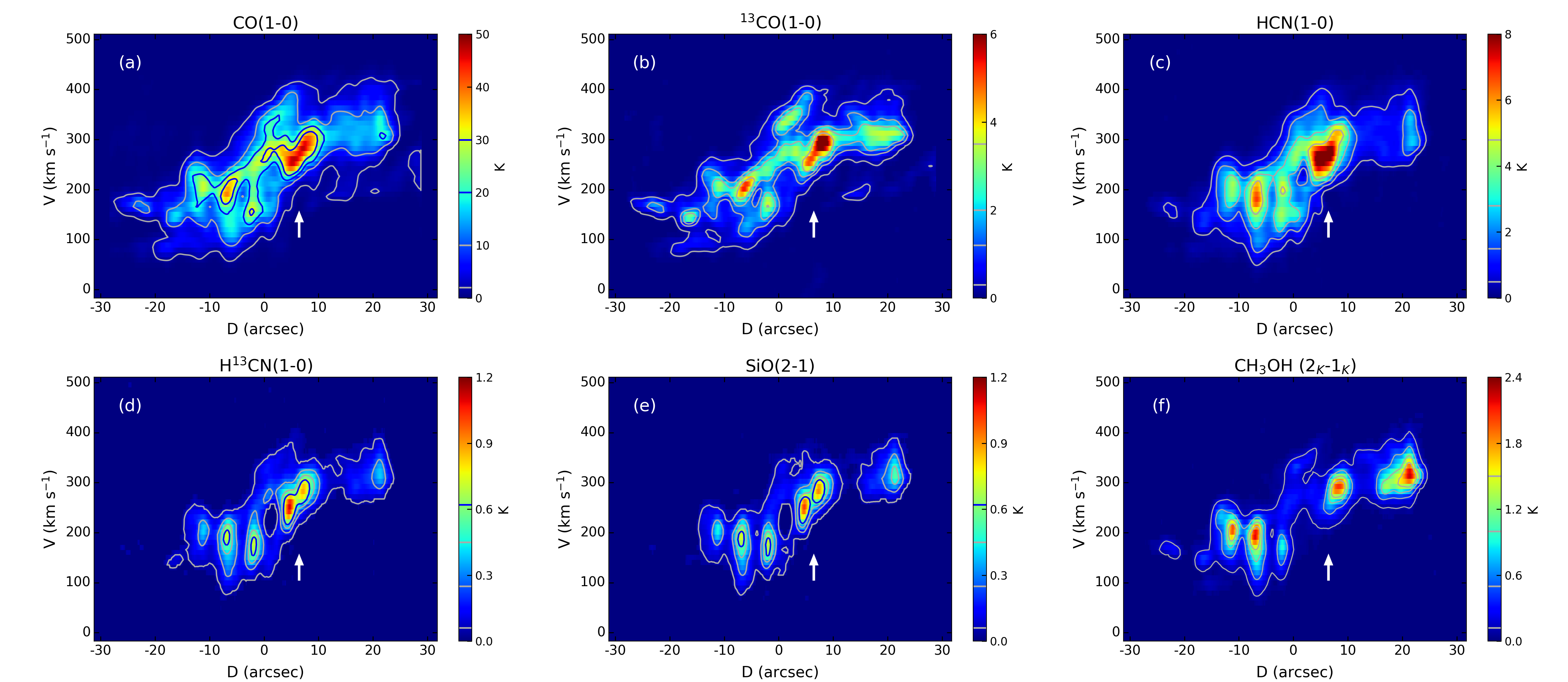}
   \caption{The intensities of molecular lines in the PVDs along the major axis. The major axis of NGC~253 is marked by the black solid line in Fig. \ref{fig1}(a), on which the black arrow shows the positive direction of position. (a) CO(1-0) PVD. The grey contours are drawn at [2, 10]\,K, and the blue contours are drawn at [20, 30]\,K. The white arrow points to the SW streamer, and remains the same in the following panels. (b) $^{13}$CO(1-0) PVD. The grey contours are drawn at [0.3, 1.2, 2.0, 3.5]\,K. (c) HCN(1-0) PVD. The grey contours are drawn at [0.3, 1.2, 2.0, 3.5]\,K. (d) H$^{13}$CN(1-0) PVD. The grey contours are drawn at [0.06, 0.25, 0.45]\,K, and the blue contour is drawn at 0.62\,K. (e) SiO(2-1) PVD. The grey contours are drawn at [0.06, 0.25, 0.45]\,K, and the blue contour is drawn at 0.62\,K. (f) CH$_{3}$OH(2$_{k}$-1$_{k}$) PVD. The grey contours are drawn at [0.12, 0.50, 1.00, 1.50]\,K.}
   \label{fig2}
\end{figure*}

Fig. \ref{fig2} shows the intensities of five molecular lines from Fig. \ref{fig1} and SiO in the J = 2-1 transition ($\sim$86.847 GHz, see \citealt{2023A&A...675A.151H} for its integrated intensity map) in the PVDs along the major axis with systemic velocity reserved and distance referenced from the phase center. The CO(1-0) PVD in Fig. \ref{fig2}(a) mainly follows a rotating pattern, which is consistent with Figure 5 from \cite{2019ApJ...881...43K}. Another common point is that the CO(1-0) PVD shows non-disk features towards both redshifted and blueshifted sides. The strong CO(1-0) emission that is marked by blue contours at [20, 30]\,K is located in the region within a galactocentric distance range of [-10, 10] arcsec and follows a rotating pattern. The white arrow in Fig. \ref{fig2}(a) points to the SW streamer, and the strongest CO(1-0) emission is located near the base of this streamer.

\begin{figure*}
   \centering
   \includegraphics[width=\hsize]{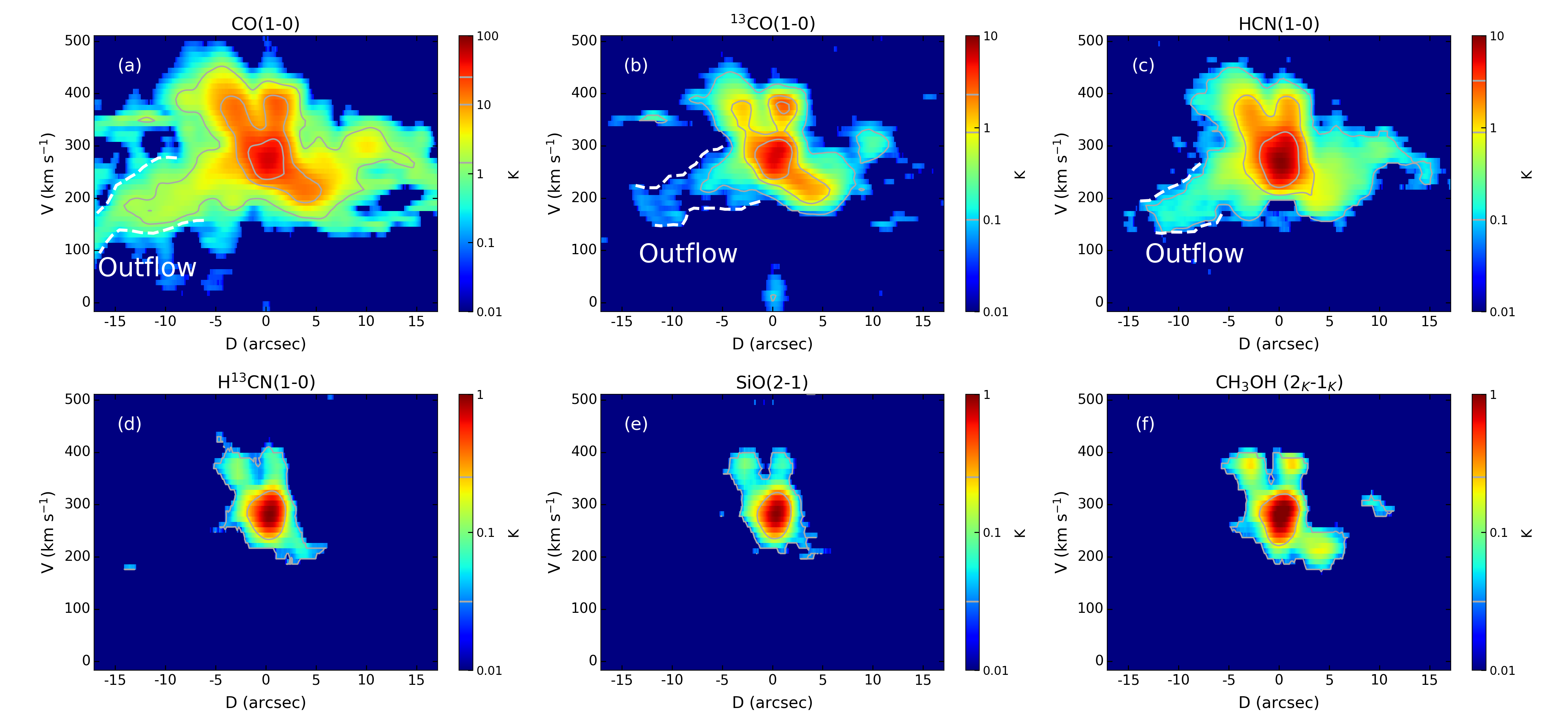}
   \caption{The intensities of molecular lines in the PVDs along the SW slice. The SW slice is marked by the black dashed line in Fig. \ref{fig1}(a), on which the black arrow shows the positive direction of the position. (a) CO(1-0) PVD. The grey contours are drawn at $\sim$[1.5, 10, 25]\,K. The white dashed profiles in panels (a), (b) and (c) outline the outflow in the SW streamer region. (b) $^{13}$CO(1-0) PVD. The grey contours are drawn at $\sim$[0.1, 0.9, 2.3]\,K. (c) HCN(1-0) PVD. The grey contours are drawn at $\sim$[0.1, 0.9, 3.2]\,K. (d) H$^{13}$CN(1-0) PVD. The grey contours are drawn at $\sim$[0.03, 0.25]\,K. (e) SiO(2-1) PVD. The grey contours are drawn at $\sim$[0.03, 0.25]\,K. (f) CH$_{3}$OH(2$_{k}$-1$_{k}$) PVD. The grey contours are drawn at $\sim$[0.03, 0.25]\,K.}
   \label{fig3}
\end{figure*}

The $^{13}$CO(1-0) PVD in Fig. \ref{fig2}(b) is less extended than the CO(1-0) and HCN(1-0) PVDs, but all have similar strong emission patterns. The H$^{13}$CN(1-0), SiO(2-1), and CH$_{3}$OH(2$_{k}$-1$_{k}$) PVDs in the second row of Fig. \ref{fig2} are the least extended. For H$^{13}$CN(1-0) and SiO(2-1) PVDs in Figs. \ref{fig2}(d) and \ref{fig2}(e), the strong emission marked by the blue contours at 0.62\,K is distributed in a few clumpy regions, among which the strongest two clumps are located near the base of the SW streamer (indicated by the white arrow). For the CH$_{3}$OH(2$_{k}$-1$_{k}$) PVD in Fig. \ref{fig2}(f), the strongest emission originates from the outskirts of the gas disk, which agrees with the integrated intensity map in Fig. \ref{fig1}(f).

Fig. \ref{fig3} shows the intensities in the PVDs along a slice across the SW streamer (SW slice, black dashed line in Fig. \ref{fig1}a) with systemic velocity reserved and distance taken zero in the slice center. The slice center is the same as Figure 1 from \cite{2017ApJ...835..265W}, and the position angle of the slice equals 150$^{\circ}$. In Fig. \ref{fig3}(a), the CO(1-0) emission in the PVD is dominated by the disk component, and is extended towards the SW streamer in the region with D$\sim$[-15, 0]\,arcsec and V$\sim$200\,km\,s$^{-1}$. Such extension outlined by the white dashed profiles, is consistent with Figure 2 from \cite{2017ApJ...835..265W}. The $^{13}$CO(1-0) and HCN(1-0) PVDs in Figs. \ref{fig3}(b) and \ref{fig3}(c) are less extended, but have similar patterns as the CO(1-0) PVD. The H$^{13}$CN(1-0), SiO(2-1) and CH$_{3}$OH(2$_{k}$-1$_{k}$) PVDs in the second row of Fig. \ref{fig3} are least extended, where the molecular emission only extends at the base of the SW streamer with D$\sim$[-5, 0]\,arcsec and V$\sim$200\,km\,s$^{-1}$. The intensities of N$_{2}$H$^{+}$(1-0) in the PVDs along the major axis and along the SW slice are also checked, which show similar patterns to the H$^{13}$CN(1-0) PVDs.

\subsection{Gas kinematics}

\begin{figure*}
   \centering
   \includegraphics[width=0.9\hsize]{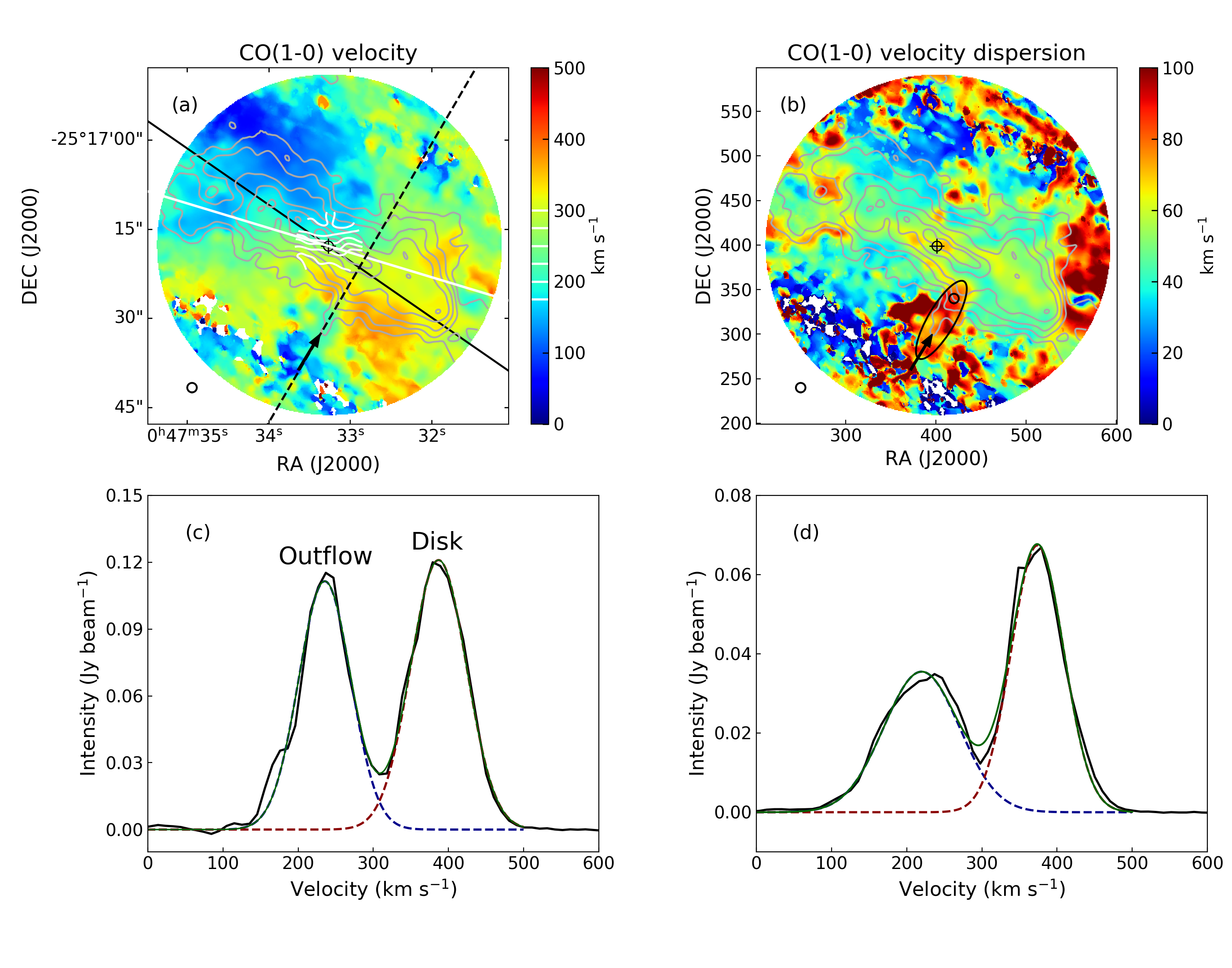}
   \caption{Kinematic features of CO(1-0). (a) Velocity field of CO(1-0). The grey contours are the same as in Fig. \ref{fig1}(a). The white contours mark the [175, 300]\,km\,s$^{-1}$ velocity range with a step of 25\,km\,s$^{-1}$ in the central $\sim$100\,pc. The white solid line shows the direction of the molecular bar. The black solid and dashed lines are the same as Fig. \ref{fig1}(a). The black arrow is the same as the white one in Fig. \ref{fig1}(a), and indicates the same region in panel (b) and the following figures. (b) Velocity dispersion field of CO(1-0). The grey contours are the same as in panel (a). The black empty ellipse outlines the high-velocity-dispersion region in the SW streamer region. The black empty circle inside the ellipse marks a beam-size region in the SW streamer region. (c) The averaged CO(1-0) spectrum from the black empty circle in panel (b). The black profile shows the CO(1-0) line. The blue dashed profile shows the blueshifted Gaussian component. The red dashed profile shows the redshifted Gaussian component. The green solid profile is the superposition. (d) The averaged CO(1-0) spectrum from the black empty ellipse in panel (b). The color codes are the same as panel (c).}
   \label{fig4}
\end{figure*}

The CO(1-0) line is generally the best tracer of total molecular gas content thanks to its high abundance and low critical density for excitation. Fig. \ref{fig4} shows the kinematics of the molecular gas in NGC~253 traced with CO(1-0). There are two interesting kinematic features in Figs. \ref{fig4}(a) and \ref{fig4}(b): (1) the velocity field showing gradients along both major and minor axes; (2) the blueshifted velocity and high velocity dispersion in the SW streamer region.

The white contours showing the velocity gradient in the center within a radius of $\sim$100\,pc in Fig. \ref{fig4}(a) have the same pattern as the CO velocity field from \cite{2001ApJ...549..896D} and \cite{2019ApJ...881...43K}. Their studies showed that the molecular bar shares the same direction as the stellar bar in the near-infrared band, and tilts $\sim$18$^{\circ}$ with respect to the major axis \citep{1996ApJ...470..821P, 2014A&A...567A..86I}. The direction of molecular and stellar bars is shown as a white solid line in Fig. \ref{fig4}(a). There is a velocity gradient (white contours) in Fig. \ref{fig4}(a) along the direction of the molecular bar (white solid line). Yet the ionized bar traced with the H92$\alpha$ recombination lines is almost parallel to the major axis \citep{2020MNRAS.493..627C}. \cite{2001ApJ...549..896D} explained this phenomenon as different velocities for the gas moving in different bar orbits and suggested that the perturbation in the gas velocity field in NGC~253 is due to an accretion event occurred $\sim$10 Myr ago.

The gas velocity in the SW streamer region (pointed out by the black arrow) is blueshifted ($\sim\rm200\,km\,s^{-1}$) as shown in Fig. \ref{fig4}(a), which means the outflowing gas is on the approaching side of NGC~253. The blueshifted outflow is visibly distinguishable from the redshifted ($\sim\rm380\,km\,s^{-1}$) disk rotation. Meanwhile, the other outflow streamers are indistinguishable in Fig. \ref{fig4}(a), because they are fainter than SW streamer and/or locate behind the galactic disk \citep{2013Natur.499..450B}. In Fig \ref{fig4}(b), the gas in the SW streamer region presents high velocity dispersion, which can be contributed by blueshifted outflow and redshifted rotation.

To compare the kinematics between the outflow and disk, we decompose two components from the CO(1-0) spectra in the SW streamer region. From a circular region in the SW streamer region having the same area as the beam size (the black empty circle in Fig. \ref{fig4}b), we extract the averaged CO(1-0) spectrum, which is displayed by the black solid profile in Fig. \ref{fig4}(c). The averaged CO(1-0) spectrum shows a double-peak structure, where we take the velocity $\sim$300 km\,s$^{-1}$ as a reference velocity for the emission near the base of the SW streamer. Combining the velocity field of CO(1-0) in Fig. \ref{fig4}(a), the blueshifted component relative to the reference velocity is emitted by the SW streamer, and the redshifted component is emitted by the gas disk. We also extract the averaged CO(1-0) spectrum from a region with high-velocity-dispersion (high-$\sigma$) in the {SW streamer region} (the black empty ellipse in Fig. \ref{fig4}b), which is displayed by the black solid profile in Fig. \ref{fig4}(d) also showing a double peak structure. Using Python-based tool \texttt{curve\_fit}, we fit the CO(1-0) spectra in Figs. \ref{fig4}(c) and \ref{fig4}(d) with double Gaussian function. The blue dashed profiles show the blueshifted Gaussian components, the red dashed profiles show the redshifted Gaussian components and the green solid profiles show the superpositions. The velocity and full width at half maximum (FWHM) in the velocity space of each Gaussian component are listed in Table \ref{tab1}.

\begin{table*}
   \caption{Results of double Gaussian fits for the CO(1-0) spectra.}             
   \label{tab1}      
   \centering          
   \begin{tabular}{c c c c c c}     
   \hline\hline       
   Region & V$_{\rm Streamer}$ (km\,s$^{-1}$) & FWHM$_{\rm Streamer}$ (km\,s$^{-1}$) & V$_{\rm Disk}$ (km\,s$^{-1}$) & FWHM$_{\rm Disk}$ (km\,s$^{-1}$) & FWHM$_{\rm Streamer}$/FWHM$_{\rm Disk}$ \\ 
   \hline                    
      Beam-size                                   &  235.5 $\pm$ 0.2 &  81.8 $\pm$ 0.6 & 386.9 $\pm$ 0.2 & 87.5 $\pm$ 0.5 & 0.93 \\
      High-$\sigma$                               &  219.7 $\pm$ 0.5 & 118.4 $\pm$ 1.3 & 373.9 $\pm$ 0.2 & 82.4 $\pm$ 0.5 & 1.44 \\
   \hline                  
   \end{tabular}
\end{table*}

We find that the blueshifted and redshifted components of the averaged CO(1-0) spectrum from the beam-size region (the black empty circle in Fig. \ref{fig4}b) have similar FWHMs in the velocity space (line widths). The redshifted component of the averaged CO(1-0) spectrum from the high-$\sigma$ region (the black empty ellipse in Fig. \ref{fig4}b) has similar FWHM to the components from the beam-size region, while the blueshifted component is $\sim$50\% wider than the redshifted component. Moreover, there is a gradual velocity shift along the SW streamer in the CO(1-0) PVD outlined by white dashed profiles in Fig. \ref{fig3}(a). The velocity gradient inside the SW streamer can be evidence for an inside-out acceleration on the gas velocity, i.e., the gas inside the SW streamer is accelerated as it outflows. It can also be attributed to the ability for the fast ejecta to reach farther away than the slow ejecta. Those two possibilities were also discussed in \cite{2017ApJ...835..265W}. If we take the velocity center of the blueshifted component from the high-$\sigma$ region ($\sim$219.7 km\,s$^{-1}$) as the averaged velocity for the SW streamer, then the projected local velocity of the SW streamer equals $\sim$80 km\,s$^{-1}$. Considering the inclination angle of the gas disk ($\sim$78.5$^{\circ}$) and the outflow being perpendicular to the disk, the deprojected local velocity of the SW streamer approaches $\sim$400 km\,s $^{-1}$, which is comparable with the outflow velocity in the ionized gas phase \citep{2011MNRAS.414.3719W}.

\subsection{Optical depth}\label{sec:opt_depth}

\begin{figure*}
   \centering
   \includegraphics[width=\hsize]{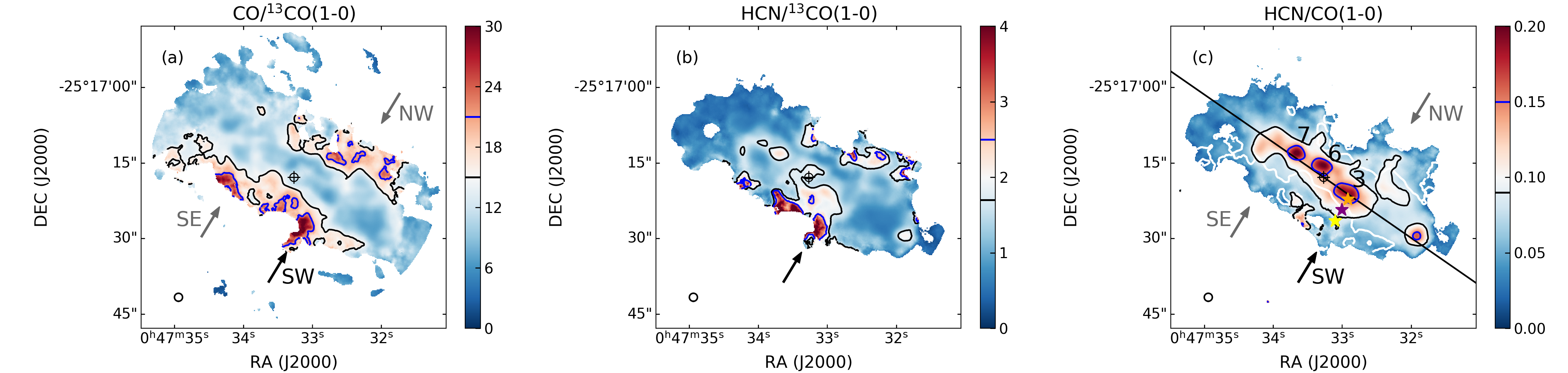}
   \caption{Integrated intensity ratio maps. (a) Integrated intensity ratio map of CO/$^{13}$CO(1-0). The ratios are taken in $\rm K\,km\,s^{-1}$ units, and remain the same in the following integrated intensity ratio plots. The black contour is drawn at 15, and the blue contour is drawn at 21. The grey arrows point to the outflow in the SE and NW directions. (b) Integrated intensity ratio map of HCN/$^{13}$CO(1-0). The black contour is drawn at 1.4, and the blue contour is drawn 2. (c) Integrated intensity ratio map of HCN/CO(1-0). The black solid line marks the major axis. The black contour is drawn at 0.09, and the blue contour is drawn at 0.15. The white contour is drawn at CO/$^{13}$CO(1-0) ratio of 15. The orange, purple, and yellow stars mark the positions at -1.5$^{\prime\prime}$, -4$^{\prime\prime}$ and -6.5$^{\prime\prime}$ offsets along the SW slice. The labels `3', `6', and `7' mark three GMCs. The grey arrows point to the outflow in the SE and NW directions.}
   \label{fig5}
\end{figure*}

Figs. \ref{fig1}(a) and \ref{fig1}(b) show the contours of CO(1-0) emission being more extended towards the SW streamer region than $^{13}$CO(1-0) emission, which can be a result of the different optical depths. Fig. \ref{fig5}(a) shows the integrated intensity ratio map of CO/$^{13}$CO(1-0) based on self-calibration, where we masked the regions with ratio S/N being less than $3\sigma$. The CO/$^{13}$CO(1-0) ratio increases in the SW, southeastern (SE) and northwestern (NW) directions (named after \citealt{2013Natur.499..450B}), which are pointed out by the black/grey arrows and are perpendicular to the gas disk (Fig.~\ref{fig5}). The black contour with CO/$^{13}$CO(1-0) ratio of 15 defines the boundary between the outflow streamers and the disk. The CO/$^{13}$CO(1-0) ratio further increases in part of the outflow region and is outlined by the blue contour at 21, where the CO/$^{13}$CO(1-0) ratio is highest in the SW streamer region. CO is generally optically thick, while its isotopologue $^{13}$CO is optically thinner. The molecular gas in the CMZ of NGC~253 shows high kinetic temperature in the previous studies \citep{2004ApJ...611..835P, 2011ApJ...735...19S}, which weakens the fractionation effects on C-bearing species \citep{2020A&A...640A..51C}. Moreover, $^{13}$C preferentially forms in the center, while the C and $^{13}$C can be well mixed in the outflow region. The CO/$^{13}$CO(1-0) ratio is expected to be similar to the C/$^{13}$C isotope ratio in the optically thin limit, and decreases as the optical depth increases.

\cite{2019A&A...624A.125M} estimated a value of the isotopic ratio C/$^{13}$C$\sim$21$\pm$6 via the integrated intensity ratio between C$^{18}$O(1-0) and $^{13}$C$^{18}$O(1-0). In Fig. \ref{fig5}(a), the observed CO/$^{13}$CO(1-0) ratio in the gas disk is lower than the C/$^{13}$C ratio, which can be contributed by the optically thick CO emission in the disk. The CO/$^{13}$CO(1-0) ratio in the SW, SE and NW outflow regions approaches or exceeds the C/$^{13}$C ratio, where the fluctuation in isotopic abundance is weak. Part of the outflow region (outlined by the blue contour at 21), including the SW streamer region, shows the CO/$^{13}$CO(1-0) ratio consistent with the C/$^{13}$C ratio. Except for NGC~253, \cite{2005A&A...438..533W} found the integrated intensity ratio CO/$^{13}$CO(1-0) of the prominent molecular streamers being comparable to that of the starburst disk in M82, which is different from the increasing CO/$^{13}$CO(1-0) ratio towards the directions of outflow in NGC~253.

\begin{figure*}
  \centering
  \includegraphics[width=0.9\hsize]{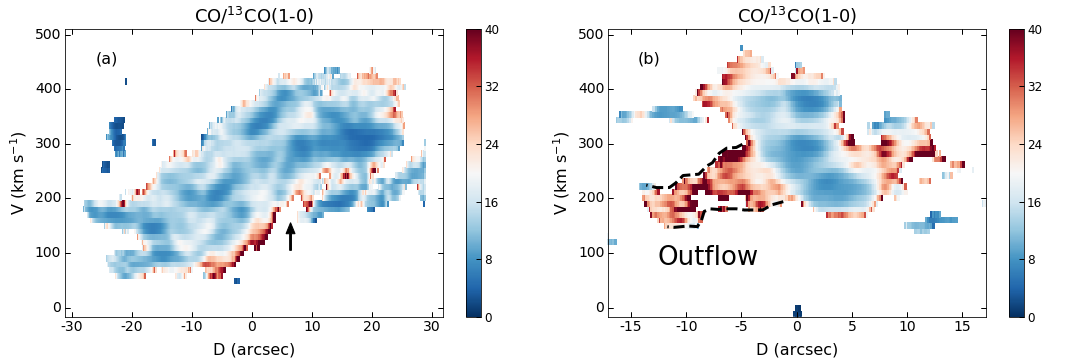}
  \caption{Intensity ratio in the PVDs. (a) Intensity ratio of CO/$^{13}$CO(1-0) in the PVD along the major axis. The black arrow points to the SW streamer, and remains the same in the following figures. (b) Intensity ratio of CO/$^{13}$CO(1-0) in the PVD along the SW slice. The black dashed profiles outline the outflow in the SW streamer region.}
  \label{fig6}
\end{figure*}

To decompose different velocity components, we plot in Figs. \ref{fig6}(a) and \ref{fig6}(b) the intensity ratio of CO/$^{13}$CO(1-0) in the PVDs along the major axis and the SW slice, which are marked by the black solid and dashed lines in Fig. \ref{fig1}(a). In Fig. \ref{fig6}(a), the CO/$^{13}$CO(1-0) ratio increases in the non-disk components and is highest on the blueshifted side, which implies the decrease in the optical depth of CO emission in the molecular gas on the approaching side of NGC~253. This phenomenon is more obvious in the PVD along the SW slice in Fig. \ref{fig6}(b), where the CO/$^{13}$CO(1-0) ratio is low along the gas disk and increases inside the SW streamer with D$\sim$[-15, 0] arcsec and V$\sim$200 km\,s$^{-1}$ outlined by the two black dashed profiles. Combining Figs. \ref{fig5}(a), \ref{fig6}(a), and \ref{fig6}(b), we infer that the decrease in the optical depth of CO emission happens inside the molecular outflow including the SW streamer, which can be attributed to the gas velocity gradient inside the outflow.

Measuring the molecular mass outflow rate is important because it may indicate the rate at which the fuel for star formation is expelled, which may significantly suppress star formation activity. The optical depth of CO emission is a key factor in estimating the molecular mass outflow rate. \cite{2018ApJ...867..111Z} derived the optical depth with the integrated intensity ratio of CO(2-1)/CO(1-0), and suggested that the majority of the CO emission is optically thick in the outflow region of NGC~253. However, the $^{13}$C-bearing molecular species being optically thinner than the $^{12}$C-bearing ones implies that the CO/$^{13}$CO(1-0) ratio is a more reliable indicator of CO optical depth. Hence, the agreement between the CO/$^{13}$CO(1-0) ratio (Fig. \ref{fig5}a) and the C/$^{13}$C ratio \citep{2019A&A...624A.125M} suggests that the CO emission in a considerable portion of the outflow region is optically thin in NGC~253. Such phenomenon supports the estimation from \cite{2013Natur.499..450B} on the total molecular mass outflow rate $\sim 9\ M_{\odot}\ yr^{-1}$, which is three times the star formation rate $\sim 3\ M_{\odot}\ yr^{-1}$ \citep{2005ApJ...629..767O} in NGC~253.

HCN is one of the most abundant high dipole-moment molecules that trace dense gas. The permanent electric dipole-moment of HCN ($\mu_{e} \sim$ 2.99 D) is much higher than CO ($\mu_{e} \sim$ 0.11 D) \citep{1984JChPh..80.3989E, 1994ApJS...95..535G}, which makes the critical density of HCN three orders of magnitude higher than that of CO. Fig. \ref{fig5}(b) shows the integrated intensity ratio map of HCN/$^{13}$CO(1-0). The HCN/$^{13}$CO(1-0) ratio is lowest in the gas disk, increases towards the outflow directions, and becomes the highest in parts of the outflow region (outlined by the blue contour at a ratio of 2) including the SW streamer region. Such a trend reminds us of the increasing CO/$^{13}$CO(1-0) ratio in the outflow region with an increasing distance from the major axis in Fig. \ref{fig5}(a). Given that HCN(1-0) emission is optically thick in the central $\sim$kpc region of NGC~253 \citep{2015ApJ...801...63M}, the increased line width attributed to gas velocity gradient in the SW streamer region not only can decrease the optical depth of CO(1-0) emission tracing molecular gas, but also can decrease the optical depth of HCN(1-0) emission tracing dense gas.

\subsection{Dense gas fraction}

The HCN/CO(1-0) ratio is widely used as an indicator of the fraction of dense gas, which is immediately responsible for the star formation inside galaxies \citep{2004ApJ...606..271G, 2004ApJS..152...63G, 2012ApJ...745..190L}. \cite{2024ApJ...961...18T} presented non-LTE analyses with ALCHEMI data for the CMZ of NGC~253, and compared the dense gas fraction in this region with the center of the Milky Way. The difference turned out to be consistent with that in the HCN/CO(1-0) ratio between the CMZ of NGC~253 and the Galactic center, which confirms the HCN/CO(1-0) ratio as a good measurement of the mass fraction of the dense gas to the entire molecular gas.

Fig. \ref{fig5}(c) shows the integrated intensity ratio map of HCN/CO(1-0). The HCN/CO(1-0) ratio is the highest in three clumpy regions along the major axis (black solid line), which are marked by the blue contour at a ratio of 0.15. The physical sizes of the three regions in Fig. \ref{fig5}(c) are $\lesssim$100 pc, which fit the scales of giant molecular clouds (GMCs), and can be further resolved into star-forming clumps ($\sim$10\,pc) \citep{2017ApJ...849...81A}. The positions of those GMCs are consistent with the observation in \cite{2015ApJ...801...25L}, which were numbered 3, 6, and 7. The bar of NGC~253 traced with ionized gas is also along the major axis \citep{2001ApJ...549..896D}. The inflowing gas along the bar interacting with the gas in the disk can trigger star formation inside those GMCs. The existence of supernova remnants and \textsc{Hii} region in the inner 200~pc of NGC~253 has been revealed about thirty years ago \citep{1997ApJ...488..621U}. The forming super star clusters and the winds from young clusters were detected in high spatial resolution \citep{2018ApJ...869..126L, 2021ApJ...912....4L, 2022ApJ...935...19L}. The GMCs 3, 6, and 7 in Fig. \ref{fig5}(c) are located at the base of outflow streamers outlined by the white contour, and the black contour at 0.09 is extended towards the directions of outflow streamers. Given that the dynamical age of the SW streamer equaling $\sim$1 Myr is short \citep{2017ApJ...835..265W}, the star formation inside the GMCs can be the engine for the outflow streamers \citep{2019ApJ...881...43K}.

\cite{2017ApJ...835..265W} measured the ratio of peak intensities between HCN(1-0) and CO(1-0), which equals $\sim$1/10 both in the SW streamer region and the starburst center of NGC~253. The orange, purple, and yellow stars in Fig. \ref{fig5}(c) mark the corresponding positions at -1.5$^{\prime\prime}$, -4$^{\prime\prime}$ and -6.5$^{\prime\prime}$ offsets from the slice center along the SW slice following \cite{2017ApJ...835..265W}. Based on the spatially resolved map, we can estimate the integrated intensity ratios at the three stars to be $\sim$0.19, 0.13, and 0.08. It is understandable that the dense gas fraction is the highest inside the GMC, and monotonously decreases away from the GMC.

\begin{figure*}
   \centering
   \includegraphics[width=\hsize]{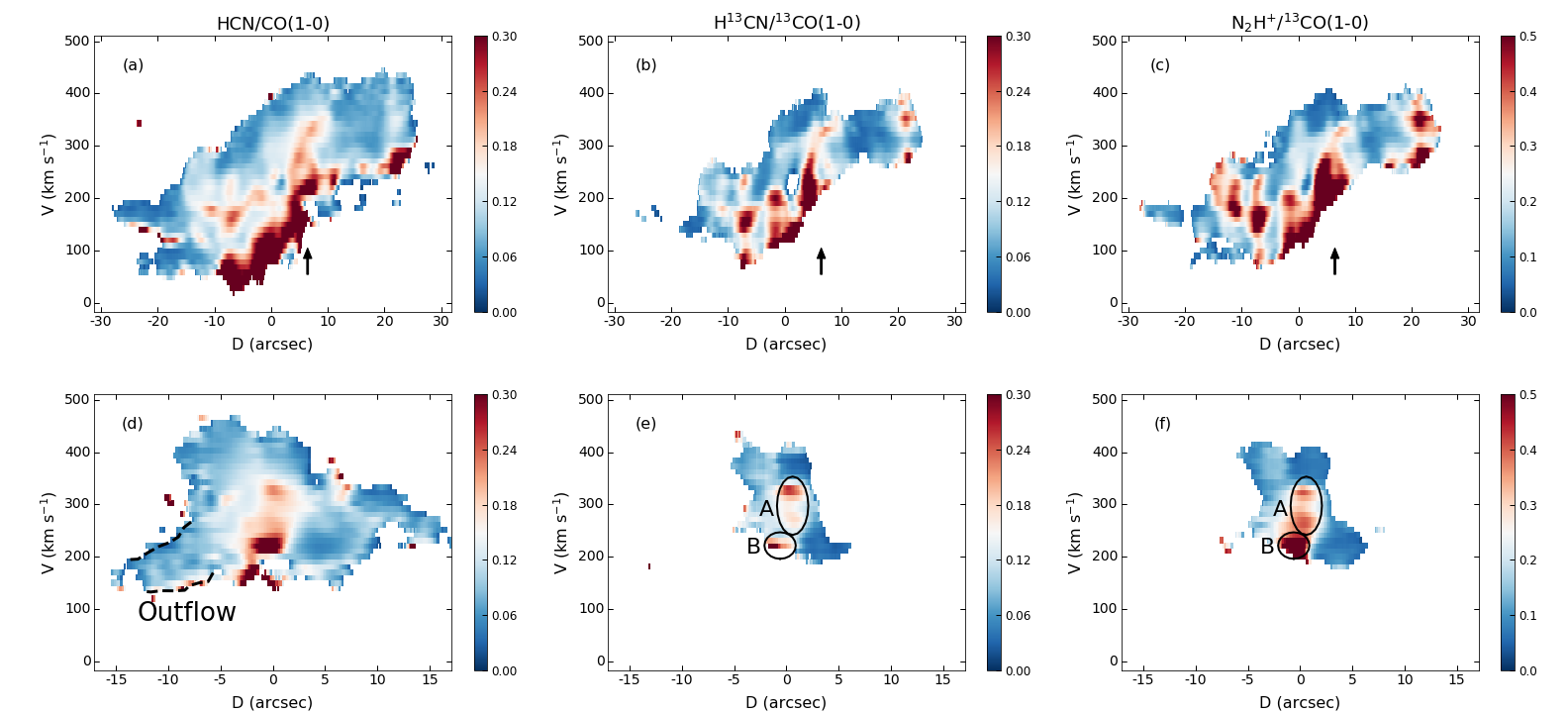}
   \caption{First row: PVDs along the major axis. Second row: PVDs along the SW slice. (a)(d) HCN/CO(1-0) ratio in PVDs. The black dashed profiles in panel (d) outline the outflow in the SW streamer region. (b)(e) H$^{13}$CN/$^{13}$CO(1-0) ratio in PVDs. In panel (e), the black ellipse named region A outlines the gas with D$\sim$0 arcsec and V$\sim$300 km\,s $^{-1}$, the black ellipse named region B outlines the gas with D$\sim$[-5, 0] arcsec and V$\sim$200\,km\,s$^{-1}$, and keep the same in the following figures. (c)(f) N$_{2}$H$^{+}$/$^{13}$CO(1-0) ratio in PVDs.}
   \label{fig7}
\end{figure*}

 Figs. \ref{fig7}(a) and \ref{fig7}(d) show the intensity ratio of HCN/CO(1-0) in the PVDs along the major axis and the SW slice. In Fig. \ref{fig7}(a), the HCN/CO(1-0) intensity ratio increases at the bottom of the PVD where the blueshifted non-disk component is located. In Fig. \ref{fig7}(d), the HCN/CO(1-0) ratio is highest in the SW streamer region with D$\sim$[-5, 0] arcsec and V$\sim$200 km\,s $^{-1}$. Figs. \ref{fig4}(c) and \ref{fig4}(d) show that the averaged projected velocity of the outflowing gas in the SW streamer region is around 200 km\,s $^{-1}$. Hence, the HCN/CO(1-0) ratio increases in the gas at the base of the SW streamer that shares the same velocity with the molecular outflow. The HCN/CO(1-0) ratio in the extended streamer region with D$\sim$[-15, -5] is moderate as in the gas disk (Fig. \ref{fig7}d), which agrees with the tendency at -6.5$^{\prime\prime}$ offset in Figure 7 from \cite{2017ApJ...835..265W}.

It is possible to detect strong molecular emission at densities below the critical density. The effective excitation density is defined as the density at which the integrated intensity of a molecular line equals 1\,K\,km\,s$^{-1}$ with reasonable assumptions about the column density and kinetic temperature. \cite{2015PASP..127..299S} calculated the optically thin critical densities and effective excitation densities at assumed column density and kinetic temperature for the common dense gas tracers. The kinetic temperatures are assumed to be $\sim$100\,K \citep{2019ApJ...871..170M} in the following comparisons on the effective excitation densities between different tracers. \cite{2015PASP..127..299S} calculated the critical density and effective excitation density of HCN(1-0) at a column density of 10$^{14}$\,cm$^{-2}$, to be $1.1 \times 10^{5}$\,cm$^{-3}$ and $1.7 \times 10^{3}$\,cm$^{-3}$, respectively.

To confirm the distribution of the dense gas in the CMZ of NGC~253, we turn to tracers of dense gas with lower opacity. The critical density and effective excitation density of H$^{13}$CN(1-0) at a column density of 10$^{12.3}$\,cm$^{-2}$, equal $9.7 \times 10^{4}$\,cm$^{-3}$ and $6.5 \times 10^{4}$\,cm$^{-3}$. Even though the higher effective excitation density originates from a lower reference column density compared with that of HCN \citep{2015PASP..127..299S}, H$^{13}$CN(1-0) as an isotopologue can trace the dense gas with less limitation of optical depth. Fig. \ref{fig7}(b) shows the intensity ratio of H$^{13}$CN/$^{13}$CO(1-0) in the PVD along the major axis, where the ratio also increases in the blueshifted non-disk component. Although the H$^{13}$CN/$^{13}$CO(1-0) ratio in the PVD along the SW slice in Fig. \ref{fig7}(e) is not as extended as HCN/CO(1-0), we still observe high ratios in two regions. One is the ellipse named region A with D$\sim$0 arcsec and V$\sim$300 km\,s $^{-1}$ (reference velocity) that represents the gas inside GMC~3. The other is the ellipse named region B with D$\sim$[-5, 0] arcsec and V$\sim$200\,km\,s$^{-1}$ (outflow velocity) that represents the gas at the base of the SW streamer.

The critical density and effective excitation density at a column density of 10$^{13}$\,cm$^{-2}$ of another dense gas tracer N$_{2}$H$^{+}$(1-0) equal $2\times10^{4}$~cm$^{-3}$ and $2.6\times10^{3}$~cm$^{-3}$. Even though the critical density of N$_{2}$H$^{+}$(1-0) is lower than HCN(1-0) and H$^{13}$CN(1-0), Galactic parsec-scale observations show that the N$_{2}$H$^{+}$(1-0) emission is exclusively associated with rather dense gas \citep{2017A&A...605L...5K, 2021A&A...646A..97T}. The intensity ratio of N$_{2}$H$^{+}$/$^{13}$CO(1-0) in the PVD along the major axis (Fig. \ref{fig7}c) has an accordant pattern with H$^{13}$CN/$^{13}$CO(1-0) PVD (Fig. \ref{fig7}b). In the PVD along the SW slice (Fig. \ref{fig7}f), the N$_{2}$H$^{+}$/$^{13}$CO(1-0) ratio increases at the base of the SW streamer (region B). On one hand, the increased patterns in Figs. \ref{fig7}(d) and \ref{fig7}(f) are highly consistent. Given that N$_{2}$H$^{+}$(1-0) is not safely optically thin and the optical depths of HCN(1-0) and CO(1-0) can be different in the central region, the N$_{2}$H$^{+}$/$^{13}$CO(1-0) and HCN/CO(1-0) ratios inside GMC~3 may be affected by optical depths. On the other hand, all three ratios in Figs. \ref{fig7}(d), \ref{fig7}(e) and \ref{fig7}(f) are higher in region B, which implies a negligible influence of optical depths at the base of the SW streamer, not to mention in the extended streamer region (Fig. \ref{fig6}b).

In summary, we find that the dense gas fraction is high inside GMC~3 (with reference velocity, region A) and at the base of the SW streamer (with outflow velocity, region B). The dense gas fraction in the extended streamer region (outlined by the black dashed profiles) is moderate in Fig. \ref{fig7}(d) and doesn't show visible signs of accumulation of dense gas traced with HCN/CO(1-0). Combining the low optical depths of CO(1-0) and HCN(1-0) emission in the extended streamer region (Figs. \ref{fig5}a and \ref{fig5}b), we suggest the existence of gas velocity gradient prevents the gas from accumulation inside the SW streamer of NGC~253.

\subsection{Shock strength}

Si-bearing material and solid-phase methanol are located in different parts of the dust grains. Silicon tends to reside in the core of grains and can be sputtered to the gas phase by fast shocks (v$_{\rm shock} \gtrsim $ 15-20 km\,s$^{-1}$, \citealt{2017A&A...597A..11K}). Gas-phase silicon reacts with molecular oxygen or a hydroxyl radical and forms SiO \citep{1997A&A...321..293S}. On the other hand, solid-phase methanol is in the icy mantle of grains. Slow shocks (v$_{\rm shock} \lesssim $ 15-20 km\,s$^{-1}$, \citealt{2022MNRAS.509.4555N}) are able to impact the icy mantle and inject gas-phase methanol into the ISM without destroying the grain core nor the methanol molecule \citep{1991ApJ...369..147M, 1995ApJ...448..232C}. \cite{2023A&A...675A.151H} studied the shock tracers SiO and HNCO with the ALCHEMI data, and revealed a picture that most of the GMCs are subjected to shocks. They declared that HNCO may not be a unique tracer of slow shocks, while a high abundance of silicon in the gas phase can only be explained by fast shocks. In this section, we will explore the methanol emission to verify the state of slow shocks, and inspect the relation between fast shocks traced by SiO emission and outflow streamers.

\begin{figure*}
   \centering
   \includegraphics[width=0.9\hsize]{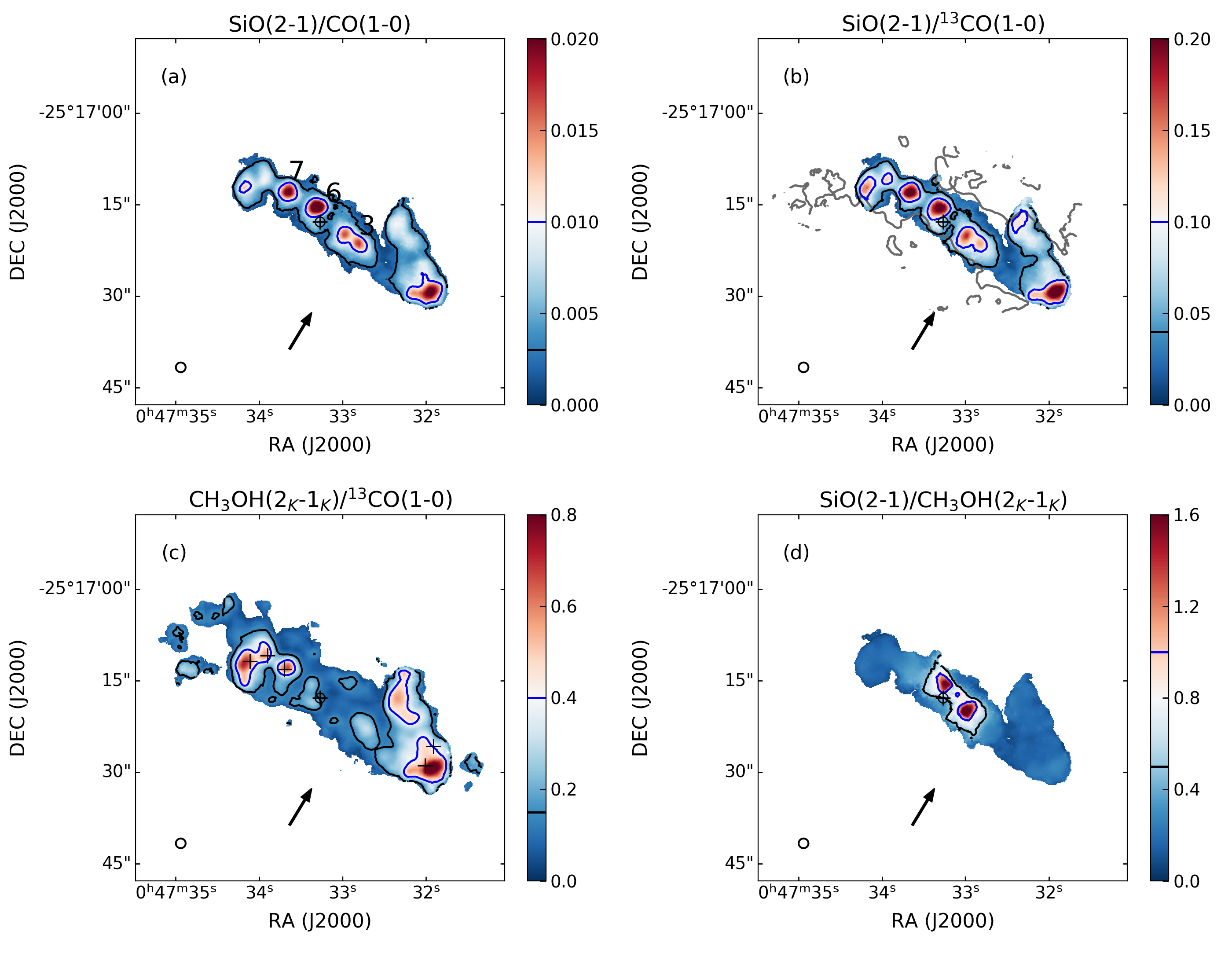}
   \caption{Integrated intensity ratio maps. (a) Integrated intensity ratio map of SiO(2-1)/CO(1-0). The black contour is drawn at 0.003, and the blue contour is drawn at 0.01. The labels `3', `6', and `7' mark the three GMCs. (b) Integrated intensity ratio map of SiO(2-1)/$^{13}$CO(1-0). The black contour is drawn at 0.04, and the blue contour is drawn at 0.1. The grey contour is drawn at CO/$^{13}$CO(1-0) ratio of 15. (c) Integrated intensity ratio map of CH$_{3}$OH(2$_{k}$-1$_{k}$)/$^{13}$CO(1-0). The black contour is drawn at 0.15, and the blue contour is drawn at 0.4. (d) Integrated intensity ratio map of SiO(2-1)/CH$_{3}$OH(2$_{k}$-1$_{k}$). The black contour is drawn at 0.5, and the blue contour is drawn at 1.}
   \label{fig8}
\end{figure*}

Fig. \ref{fig8}(a) shows the integrated intensity ratio map of SiO(2-1)/CO(1-0). The three dense GMCs (labeled by `3', `6', `7' following Fig. \ref{fig5}c) also show increased SiO(2-1)/CO(1-0) ratio that is marked by the blue contour at 0.01. Moreover, two SiO(2-1)/CO(1-0) ratio peaks appear around GMC~3 in Fig. \ref{fig8}(a), with the second one consistent with the position of GMC~4 in \cite{2015ApJ...801...25L}. To examine the influence of optical depth, we also plot the integrated intensity ratio map of SiO(2-1)/$^{13}$CO(1-0) in Fig. \ref{fig8}(b). The increased SiO(2-1)/$^{13}$CO(1-0) ratio exists in the four GMCs (marked by blue contour at 0.1). The increased ratios in the GMCs imply the enhanced fast shocks there. The grey contour in Fig. \ref{fig8}(b) outlines the outflow streamers that originate from the four GMCs. Both the distributions of SiO(2-1)/CO(1-0) (marked by black contour at 0.003) and SiO(2-1)/$^{13}$CO(1-0) (marked by black contour at 0.04) are extended towards the outflow streamers, which connect the fast shocks with the formation of the molecular outflow.

\begin{figure*}
   \centering
   \includegraphics[width=\hsize]{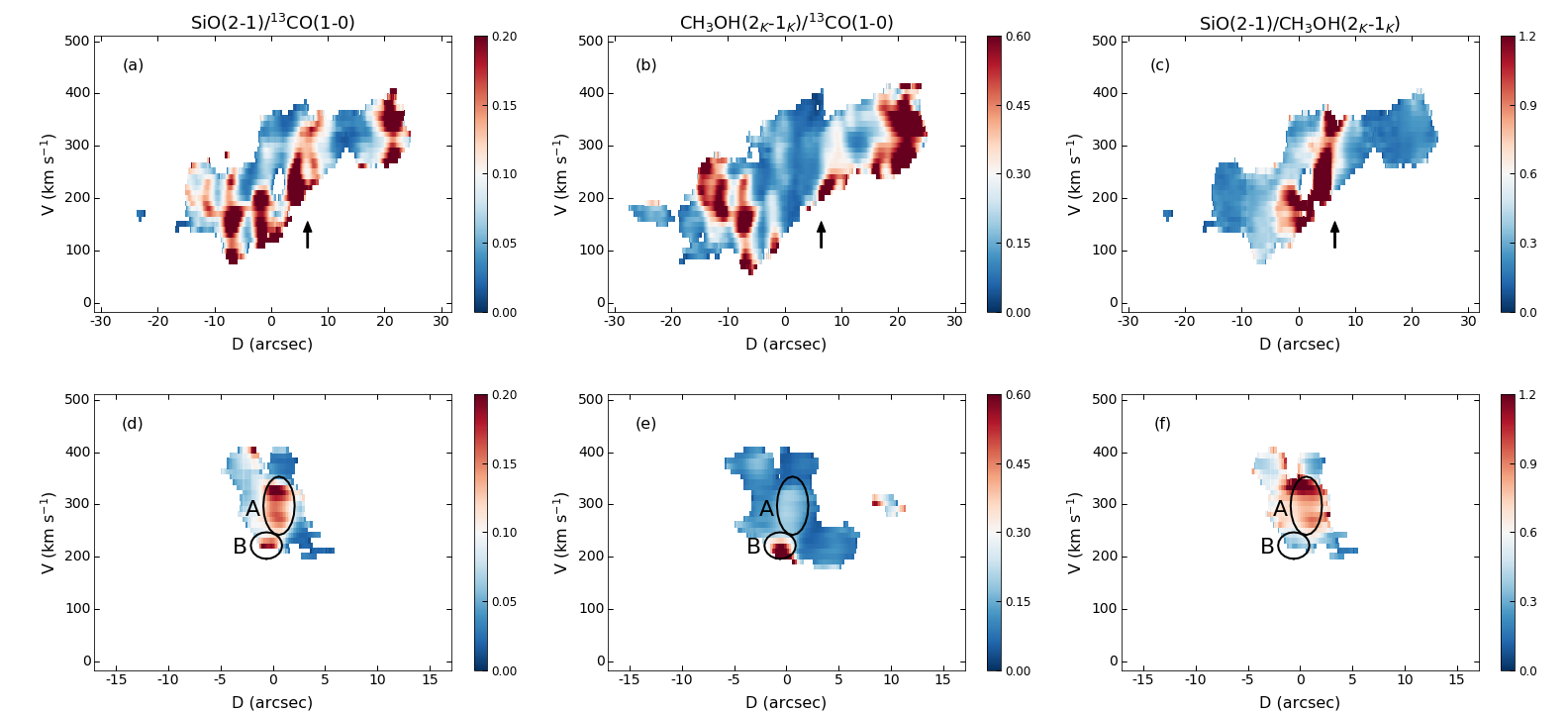}
   \caption{First row: PVDs along the major axis. Second row: PVDs along the SW slice. (a)(d) SiO(2-1)/$^{13}$CO(1-0) ratio in PVDs. (b)(e) CH$_{3}$OH(2$_{k}$-1$_{k}$)/$^{13}$CO(1-0)  ratio in PVDs. (c)(f) SiO(2-1)/CH$_{3}$OH(2$_{k}$-1$_{k}$) ratio in PVDs.}
   \label{fig9}
\end{figure*}

Fig. \ref{fig8}(c) shows the integrated intensity ratio map of CH$_{3}$OH(2$_{k}$-1$_{k}$)/$^{13}$CO(1-0). The increased CH$_{3}$OH(2$_{k}$-1$_{k}$)/$^{13}$CO(1-0) ratio that is marked by black contour at 0.15 located on the outskirts of the gas disk. Locations with high CH$_{3}$OH(2$_{k}$-1$_{k}$)/$^{13}$CO(1-0) ratio are similar to those with high CH$_{3}$OH(2$_{k}$-1$_{k}$) integrated intensity shown in Fig. \ref{fig1}(f). Such a tendency is in agreement with HNCO 4$_{0,4}$-3$_{0,3}$ emission in \cite{2023A&A...675A.151H} that shows high integrated intensity in the outermost CMZ of NGC~253. The strong methanol emission on the outskirts, with positions in consistent with the methanol masers found by \cite{2017ApJ...842..124G}, implies the frequent occurrence of slow shocks. Meanwhile, the weak methanol emission in the center could be a result of the infrequent slow shocks, or the depletion of methanol molecules \citep{1995MNRAS.272..184H, 2017MNRAS.472..604E} by fast shocks or photo-dissociation due to intense star formation \citep{2022A&A...663A..33H}. The non-cospatial distributions of fast and slow shocks were also found in another nearby galaxy NGC 1068 \citep{2017A&A...597A..11K} and imply that fast shocks do not necessarily occur with slow shocks. To obtain the relative strength between the fast and slow shocks, we plot the integrated intensity ratio map of SiO(2-1)/CH$_{3}$OH(2$_{k}$-1$_{k}$) in Fig. \ref{fig8}(d), where the central region turns out to be dominated by the fast shocks.

Fig. \ref{fig9} shows intensity ratios of the shock tracers in PVDs, where the top panels show the PVDs along the major axis and the bottom panels show the PVDs along the SW slice. The asymmetric and increased pattern (including regions A and B) of SiO(2-1)/$^{13}$CO(1-0) ratio in PVDs (Figs. \ref{fig9}a and \ref{fig9}d) are highly similar to the H$^{13}$CN/$^{13}$CO(1-0) ratio in PVDs (Figs. \ref{fig7}b and \ref{fig7}e). We will study in detail the correlation between the dense gas fraction and the strength of fast shocks in the next section. The increased intensity ratio of CH$_{3}$OH(2$_{k}$-1$_{k}$)/$^{13}$CO(1-0) on the outskirts of the gas disk in Fig. \ref{fig9}(b) is coherent with the distribution of the increased integrated intensity ratio in Fig. \ref{fig8}(c). It is worth noting that a few red pixels located in region B of Fig. \ref{fig9}(e) indicate an enhancement of the slow shocks at the base of the SW streamer. In Figs. \ref{fig9}(c) and \ref{fig9}(f), we plot the SiO(2-1)/CH$_{3}$OH(2$_{k}$-1$_{k}$) ratio in PVDs. The region A in Fig. \ref{fig9}(f) is dominated by the fast shocks, while the SiO(2-1)/CH$_{3}$OH(2$_{k}$-1$_{k}$) ratio in region B is not so high as region A. We infer that the fast shocks are triggered by the star formation inside GMC, which supports the second scenario for the origin of shocks in \cite{2023A&A...675A.151H}. Meanwhile, the fast and slow shocks co-exist in the SW streamer, and can further be related to the formation of the molecular outflow.

\section{Formation of the molecular outflow}
\label{formation of molecular outflow}

\begin{figure*}
   \centering
   \includegraphics[width = 0.9\hsize]{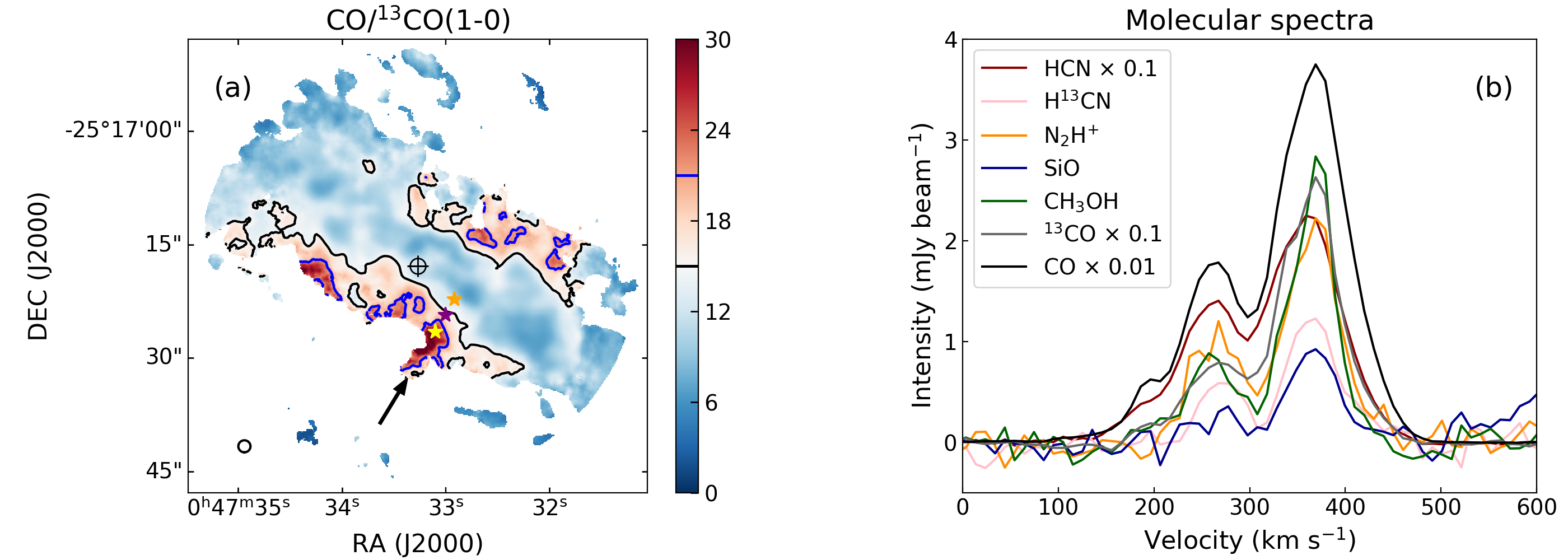}
   \caption{Molecular spectra in the SW streamer region. (a) Integrated intensity ratio map of CO/$^{13}$CO(1-0). The black contour is drawn at 15, and the blue contour is drawn at 21. The orange, purple, and yellow stars are the same as Fig. \ref{fig5}(c). (b) Averaged molecular spectra extracted from a beam size region centering at the purple star in panel (a). The color codes are labeled in the top left. The averaged intensity of the CO(1-0) line is scaled down by a factor of 100. The averaged intensities of $^{13}$CO(1-0) and HCN(1-0) lines are scaled down by a factor of 10.}
   \label{fig10}
   \end{figure*}

\cite{2017ApJ...835..265W} presented molecular spectra in the SW streamer region and quantified the relationship between the different molecular transitions, including CO(1-0) and HCN(1-0). Based on the data from the ALCHEMI survey, we can add several weaker molecular lines. In Fig. \ref{fig10}(a), we plot orange, purple, and yellow stars the same as Fig. \ref{fig5}(c) onto the self-calibrated integrated intensity ratio map of CO/$^{13}$CO(1-0), where the increased ratio indicated by red color represents the outflow region. The purple star is located at -4$^{\prime\prime}$ offset on the SW slice following \cite{2017ApJ...835..265W}. The molecular emission shown in Fig. \ref{fig10}(b) are averaged intensity profiles from a beam-size region centering on this purple star. The black profile shows the averaged intensity of the CO(1-0) spectrum (scaled down by a factor of 100), the red and grey profiles show the averaged intensities of HCN(1-0) and $^{13}$CO(1-0) spectra (scaled down by a factor of 10), as well as pink, orange, blue and green profiles, show the averaged intensities of H$^{13}$CN(1-0), N$_{2}$H$^{+}$(1-0), SiO(2-1) and CH$_{3}$OH(2$_{k}$-1$_{k}$) spectra.

\begin{table*}
   \caption{Peak intensities from double Gaussian fits for molecular spectra from the SW streamer region (Fig. \ref{fig10}b).}
   \label{tab2}      
   \centering
   \begin{threeparttable}
   \begin{tabular}{c c c c}     
   \hline\hline       
   Line & Streamer (mJy beam$^{-1}$) & Disk (mJy beam$^{-1}$) & Streamer-to-disk ratio\\ 
   \hline                    
      CO(1-0)                             & 161.75 $\pm$ 1.43 & 367.44 $\pm$ 1.47 & 0.440 $\pm$ 0.004 \\
      $^{13}$CO(1-0)                      &   7.54 $\pm$ 0.36 &  25.17 $\pm$ 0.38 & 0.300 $\pm$ 0.015 \\
      HCN(1-0)*                           &  12.60 $\pm$ 1.29 &  22.02 $\pm$ 1.43 & 0.572 $\pm$ 0.069 \\
      H$^{13}$CN(1-0)*                    &   0.64 $\pm$ 0.10 &   1.22 $\pm$ 0.09 & 0.524 $\pm$ 0.091 \\  
      N$_{2}$H$^{+}$(1-0)*                &   1.07 $\pm$ 0.05 &   2.14 $\pm$ 0.05 & 0.500 $\pm$ 0.028 \\
      SiO(2-1)                            &   0.27 $\pm$ 0.42 &   0.93 $\pm$ 0.41 & 0.290 $\pm$ 0.471 \\
      CH$_{3}$OH(2$_{k}$-1$_{k}$)         &   0.76 $\pm$ 0.09 &   2.66 $\pm$ 0.10 & 0.286 $\pm$ 0.034 \\
   \hline
   \end{tabular}
    \begin{tablenotes}
        \footnotesize
        \item[*] The star symbols in the first column mark the dense gas tracers.
    \end{tablenotes}
    \end{threeparttable}
\end{table*}

All the molecular lines in Fig. \ref{fig10}(b) show double peak structures, which are in agreement with Figure 7 from \cite{2017ApJ...835..265W} and CO(3-2) lines in Figure 4 from \cite{2022ApJ...935...19L}. The relatively weak blueshifted component is emitted by the SW streamer, and the relatively strong redshifted component is emitted by the gas disk. We use the Python-based tool \texttt{curve\_fit} to conduct a double Gaussian fit on each molecular line. The peak intensities together with errors from the Gaussian fit of the streamer and disk components for each molecular line are listed in the second and third columns of Table \ref{tab2}, and the corresponding line name is listed in the first column. Although the streamer component is weak in the SiO(2-1) line, it exists in all the molecular lines that are extracted from the beam-size region (purple star in Fig. \ref{fig10}a) in the SW streamer.

As described in the introduction, there are three main formation scenarios previously proposed for the molecular outflow. One is the molecular outflow directly driven by radiation and/or pressure, one is the molecular cloud entrained by the hot wind, and the other is the molecular outflow in situ forming from the hot wind. The last scenario assumes the cooling timescale to be shorter than the dynamical timescale \citep{2000MNRAS.317..697E, 2003ApJ...590..791S}. \cite{2017ApJ...835..265W} estimated that the molecular gas inside the SW streamer of NGC~253 is ejected from the disk $\sim$1 Myr ago, which is too short to create new CO molecules that can emit \citep{2012MNRAS.424.2599C}. Moreover, the phenomenon that all the molecular lines in Table \ref{tab2} show streamer features indicates that the SW streamer starts as dense, shocked, and chemically rich outflowing gas. The existence of dense gas tracers, e.g., HCN, H$^{13}$CN and N$_{2}$H$^{+}$, inside the outflow further excludes the possibility of in situ formation. However, it is still difficult to distinguish whether the outflowing molecular gas is directly driven by radiation/pressure or is entrained by the hot wind.

\subsection{Correlation between the dense gas fraction and the strength of fast shocks}

\begin{figure*}
   \centering
   \includegraphics[width = 0.9\hsize]{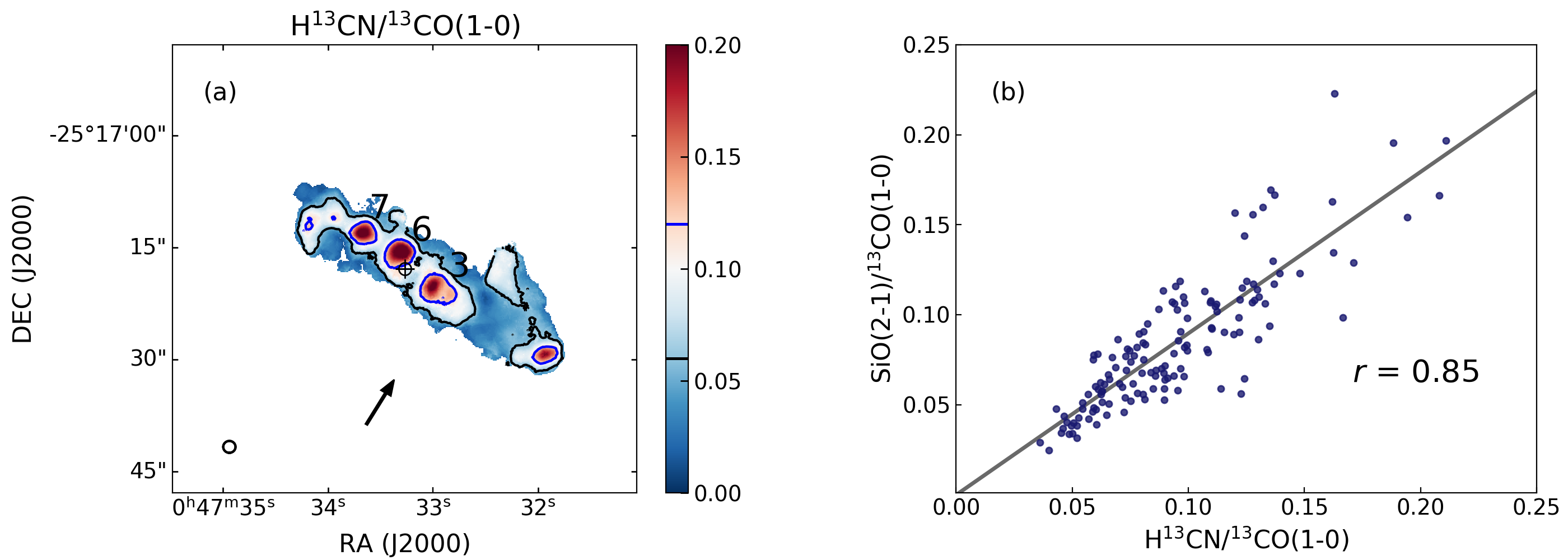}
  \caption{Correlation between the dense gas fraction and the strength of fast shocks. (a) Integrated intensity ratio map of H$^{13}$CN/$^{13}$CO(1-0). The black contour is drawn at 0.06, and the blue contour is drawn at 0.12. The labels `3', `6', and `7' mark the same GMCs as Fig. 5(c). (b) Correlation between H$^{13}$CN/$^{13}$CO(1-0) and SiO(2-1)/$^{13}$CO(1-0). The blue points mark the ratios from pixels with a spatial resolution equaling the beam size. The grey solid line shows the result of a linear fit to the blue points.}
   \label{fig11}
\end{figure*}

Fig. \ref{fig11}(a) shows the integrated intensity ratio map of H$^{13}$CN/$^{13}$CO(1-0), where the GMCs have the highest ratios marked by the blue contour at 0.12, and the black contour at 0.06 is extended towards the outflow streamers. Such increased patterns are similar to that of SiO(2-1)/$^{13}$CO(1-0) in Fig. \ref{fig8}(b). Moreover, the increased H$^{13}$CN/$^{13}$CO(1-0) and SiO(2-1)/$^{13}$CO(1-0) ratios in PVDs in Figs. \ref{fig7} and \ref{fig9} also share similar patterns. The consistency in the enhanced emissions of H$^{13}$CN(1-0) and SiO(2-1) implies a positive correlation between the dense gas fraction and the strength of fast shocks in NGC~253. We regrid the H$^{13}$CN(1-0), SiO(2-1) and $^{13}$CO(1-0) data cubes to a 1.6$^{\prime\prime}$ beam-size to avoid oversampling, then statistically quantify the linear relationship between H$^{13}$CN/$^{13}$CO(1-0) and SiO(2-1)/$^{13}$CO(1-0).

The blue points in Fig. \ref{fig11}(b) show the H$^{13}$CN/$^{13}$CO(1-0) and SiO(2-1)/$^{13}$CO(1-0) ratios in each beam size region from Figs \ref{fig8}(b) and \ref{fig11}(a). To check the linear relationship between two ratios, which is visibly tight, we calculate the Pearson correlation coefficient using the Python-based package \texttt{pearsonr}. The correlation coefficient ($r$) equals 0.85, which reveals a tight positive correlation between H$^{13}$CN/$^{13}$CO(1-0) and SiO(2-1)/$^{13}$CO(1-0) ratios. We perform a linear fit for the data set, which is shown by the grey solid line in Fig. \ref{fig11}(b). We also calculate the Pearson correlation coefficient between H$^{13}$CN(1-0) and SiO(2-1) integrated intensities, which approaches 0.96 and reveals a tight correlation between the intensities of two lines. Therefore, the correlation between H$^{13}$CN/$^{13}$CO(1-0) and SiO(2-1)/$^{13}$CO(1-0) ratios is real, it is not a result from the common denominator. Meanwhile, we keep in mind that the high critical densities of H$^{13}$CN(1-0) and SiO(2-1) may contribute to the tight positive correlation between two ratios. Theoretically, the star formation that happens in the region with a high dense gas fraction can trigger fast shocks. Combining the similar extensions of the SiO(2-1) and H$^{13}$CN(1-0) emissions towards the outflow in Figs \ref{fig8}(b) and \ref{fig11}(a), the star formation inside the GMCs can trigger fast shocks and contribute to the formation of the molecular outflow in NGC~253.

The difference between the tracers of dense gas and shocks presents when we further analyse the molecular spectra in Fig. \ref{fig10}(b). Given that all the lines show double peak structures, we can distinguish the contributions of steamer and disk via the double Gaussian fits. In the last column of Table \ref{tab2}, we list the streamer-to-disk peak intensity ratios (streamer-to-disk ratios for short) together with errors for different molecular lines. Each ratio is calculated by dividing the peak intensity of the streamer component by that of the disk component.

The streamer-to-disk ratio of CO(1-0) is higher than its isotopologue $^{13}$CO(1-0), which can be explained by the lower opacity environment inside the SW streamer than in the gas disk for CO emission as discussed in Section \ref{sec:opt_depth}. It is interesting to find that all the dense gas tracers (marked by the star symbols in the first column of Table \ref{tab2}) have higher streamer-to-disk ratios than the others. The optical depth decreasing inside the SW streamer for HCN emission can explain the highest streamer-to-disk ratio of HCN(1-0) among the dense gas tracers. In addition, the SiO(2-1) and methanol, as the tracers of fast and slow shocks, respectively, have similar streamer-to-disk ratios to $^{13}$CO(1-0). It seems that the enhancement of dense gas fraction is stronger in the SW streamer than in the gas disk, while the shock strength is equivalently enhanced in the SW streamer and the gas disk. More detailed clues are needed to explain such a difference.

We suggest the physical pictures that are related to the SW streamer in NGC~253 as follows: (i) The GMCs (Fig. \ref{fig5}c) in the direction of the major axis are related to the gas accretion along the bar structure that further provides material for the star formation. Meanwhile, the star formation inside the GMCs that are located at the base of the outflow streamers contributes to driving the molecular outflow, which is in agreement with the model presented by \cite{2022ApJ...935...19L}. (ii) There can be intense star formation inside the GMCs also presenting as higher dense gas fraction, and results in enhanced shock strength at the base of the outflow (Figs. \ref{fig7} and \ref{fig9}) including the SW streamer. (iii) The optical depths of CO and HCN emission decrease in the SW streamer (Figs. \ref{fig5}a and \ref{fig5}b), and the dense gas is diluted in the extended SW streamer region (Fig. \ref{fig7}d), which can be attributed to the gas velocity gradient inside the molecular outflow.

\section{Summary}
\label{summary}

In this work, we analyse data from the ALCHEMI survey and study the physical properties of the molecular outflow in the starburst galaxy NGC~253.

\begin{itemize}
   \item [(1)] The emission of CO(1-0), $^{13}$CO(1-0), HCN(1-0), H$^{13}$CN(1-0), N$_{2}$H$^{+}$(1-0) and CH$_{3}$OH(2$_{k}$-1$_{k}$) is extended towards the SW streamer. The CO(1-0) and HCN(1-0) emission is the most extended, which can be attributed to the optically thick environments in the gas disk and decreased optical depths towards the SW streamer.
   \item [(2)] The molecular outflow in the SW streamer region is blueshifted with a deprojected local velocity of $\sim$400 km\,s$^{-1}$, which is consistent with previous studies. The wider blueshifted component than the disk component suggests an inside-out acceleration of molecular outflow or the fast ejecta getting farther away than the slow ejecta. All the molecular spectra from the SW streamer region show double peak structures, which indicate that the SW streamer starts as dense, shocked and chemically rich outflowing gas, rather than in situ formation from hot wind.
   \item [(3)] The integrated intensity ratio maps of CO/$^{13}$CO(1-0) and HCN/$^{13}$CO(1-0) show similar patterns. Both are lowest in the gas disk, increase towards outflow directions that are perpendicular to the gas disk, and become highest in the SW streamer region. Combining the isotopic ratio of C/$^{13}$C from a previous study, we suggest that the CO(1-0) emission is optically thin in the SW streamer region, which may be a result of the gas velocity gradient.
   \item [(4)] Three GMCs present in the integrated intensity ratio map of HCN/CO(1-0), which are aligned with the direction of the major axis and can be caused by the gas accretion along the bar structure. Those GMCs are located at the base of the outflow streamers including the SW streamer, where the star formation can drive molecular outflow. The HCN/CO(1-0), H$^{13}$CN/$^{13}$CO(1-0) and N$_{2}$H$^{+}$/$^{13}$CO(1-0) intensity ratios in PVDs show high dense gas fraction at the base of the SW streamer. The HCN/CO(1-0) intensity ratio in PVDs suggests moderate dense gas fraction in the extended streamer region without visible signs of accumulation of dense gas, which may also be a result of gas velocity gradient.
   \item [(5)] The SiO(2-1)/CO(1-0) and SiO(2-1)/$^{13}$CO(1-0) integrated intensity ratios show enhanced fast shocks in the GMCs. The SiO(2-1)/$^{13}$CO(1-0) and CH$_{3}$OH(2$_{k}$-1$_{k}$)/$^{13}$CO(1-0) intensity ratios in PVDs indicate that fast shocks can be triggered by the star formation inside GMC, while fast and slow shocks co-exist at the base of the SW streamer and can be related to the formation of molecular outflow.
   \item [(6)] There is a tight positive correlation between the dense gas fraction traced with H$^{13}$CN/$^{13}$CO(1-0) and the strength of fast shocks traced with SiO(2-1)/$^{13}$CO(1-0). The dense gas fraction is high in GMCs, where the star formation can trigger fast shocks and contribute to the formation of molecular outflow. One difference is that the enhancement of the dense gas fraction is more tightly related to the SW streamer than the gas disk, while the shock strength is equivalently enhanced in the SW streamer and the gas disk.
\end{itemize}

\begin{acknowledgements}
We thank the referee for constructive comments that greatly improved this paper. This paper makes use of the following ALMA data: ADS/JAO.ALMA\#2017.1.00161.L. ALMA is a partnership of ESO (representing its member states), NSF (USA) and NINS (Japan), together with NRC (Canada), MOST and ASIAA (Taiwan), and KASI (Republic of Korea), in cooperation with the Republic of Chile. The Joint ALMA Observatory is operated by ESO, AUI/NRAO and NAOJ.
This work was supported by JSPS KAKENHI Grant Number JP17H06130 and the NAOJ ALMA Scientific Research Grant Number 2017-06B. M.B. acknowledges support from the National Natural Science Foundation of China (No. 12303009) and the China Scholarship Council (No. 202006860042). N.H. acknowledges support from JSPS KAKENHI Grant Number JP21K03634. F.E. is financially supported by JSPS KAKENHI Grant Numbers JP17K14259 and JP20H00172. L.C. acknowledges financial support through the Spanish grant PID2019-105552RB-C41 funded by MCIN/AEI/10.13039/501100011033. V.M.R. has received support from the project RYC2020-029387-I funded by MCIN/AEI /10.13039/501100011033. S.V., M.B., and K-Y. Huang acknowledge support from the European Research Council (ERC) under the European Union’s Horizon 2020 research and innovation programme MOPPEX 833460. M.B. acknowledges Prof. Yu Gao for his remarkable discovery of the HCN J = 1-0 as an indicator of dense molecular gas, as well as for his gentle clarification to her regarding the ALMA instrument in 2018.

\end{acknowledgements}

\end{document}